\documentstyle[12pt]{article}

\font\twlgot =eufm10 scaled \magstep1 \font\egtgot =eufm8
\font\sevgot =eufm7 \font\twlmsb =msbm10 scaled \magstep1
\font\egtmsb =msbm8 \font\sevmsb =msbm7

\newfam\gotfam
\def\pgot{\fam\gotfam\twlgot}
\textfont\gotfam\twlgot \scriptfont\gotfam\egtgot
\scriptscriptfont\gotfam\sevgot
\def\got{\protect\pgot}
\newfam\msbfam
\textfont\msbfam\twlmsb \scriptfont\msbfam\egtmsb
\scriptscriptfont\msbfam\sevmsb
\def\Bbb{\protect\pBbb}
\def\pBbb{\relax\ifmmode\expandafter\Bb\else\typeout{You cann't use
Bbb in text mode}\fi}
\def\Bb #1{{\fam\msbfam\relax#1}}

\newcommand{\gd}{{\got d}}
\newcommand{\ccG}{{\got g}}
\newcommand{\gA}{{\got A}}

\def\thebibliography#1{\bigskip\section*{}\bigskip\list
{$^{\arabic{enumi}}$}{\settowidth\labelwidth{#1}\leftmargin\labelwidth
\advance\leftmargin\labelsep
\usecounter{enumi}}
\def\newblock{\hskip .11em plus .33em minus .07em}
\sloppy\clubpenalty4000\widowpenalty4000 \sfcode`\.=1000\relax}

\def\op#1{\mathop{\fam0 #1}\limits}

\newcommand{\nm}[1]{\mid {#1}\mid}
\newcommand{\beq}{\begin{equation}}
\newcommand{\eeq}{\end{equation}}
\newcommand{\ben}{\begin{eqnarray}}
\newcommand{\een}{\end{eqnarray}}
\newcommand{\be}{\begin{eqnarray*}}
\newcommand{\ee}{\end{eqnarray*}}
\newcommand{\bea}{\begin{eqalph}}
\newcommand{\eea}{\end{eqalph}}
\newcommand{\cA}{{\cal A}}

\newcommand{\cP}{{\cal P}}

\newcommand{\cV}{{\cal V}}
\newcommand{\cC}{{\cal C}}
\newcommand{\cL}{{\cal L}}
\newcommand{\cE}{{\cal E}}

\newcommand{\cS}{{\cal S}}
\newcommand{\cO}{{\cal O}}
\newcommand{\bL}{{\bf L}}

\newcommand{\bE}{{\bf E}}

\newcommand{\al}{\alpha}
\newcommand{\vr}{\varrho}

\newcommand{\dl}{\delta}
\newcommand{\la}{\lambda}
\newcommand{\La}{\Lambda}
\newcommand{\f}{\phi}
\newcommand{\om}{\omega}

\newcommand{\m}{\mu}

\newcommand{\e}{\epsilon}

\newcommand{\G}{\Gamma}
\newcommand{\th}{\theta}

\newcommand{\vt}{\vartheta}
\newcommand{\vf}{\varphi}
\newcommand{\up}{\upsilon}

\newcommand{\di}{{\rm dim\,}}
\newcommand{\rdr}{\stackrel{\leftarrow}{\dr}{}}

\newcommand{\si}{\sigma}
\newcommand{\Si}{\Sigma}
\newcommand{\w}{\wedge}
\newcommand{\wt}{\widetilde}

\newcommand{\ol}{\overline}
\newcommand{\dr}{\partial}
\newcommand{\ar}{\op\longrightarrow}
\newcommand{\ot}{\otimes}
\newcommand{\ap}{\approx}

\let\ssection=\section
\renewcommand{\section}{\setcounter{equation}{0}\ssection}

\newcounter{eqalph}
\newcounter{equationa}
\newcounter{remark}
\newcounter{example}
\newcounter{theorem}
\newcounter{proposition}
\newcounter{lemma}
\newcounter{corollary}
\newcounter{definition}
\newenvironment{eqalph}{\stepcounter{equation}
\setcounter{equationa}{\value{equation}} \setcounter{equation}{0}

\begin{eqnarray}}{\end{eqnarray}\setcounter{equation}{\value{equationa}}}

\def\theremark{\arabic{remark}}

\def\thetheorem{\arabic{theorem}}

\newenvironment{proof}{\medskip\noindent
{\it Proof:}}{\medskip}
\newenvironment{rem}{\refstepcounter{remark}\medskip\noindent{\it
Remark \theremark:}}{\medskip}
\newenvironment{ex}{\refstepcounter{remark}\medskip\noindent{\it Example
\theremark:}}{\medskip}
\newenvironment{theo}{\refstepcounter{theorem} \medskip
\noindent{\it Theorem \thetheorem:}}{\medskip}
\newenvironment{prop}{\refstepcounter{theorem} \medskip
\noindent{\it Proposition \thetheorem:}}{\medskip}

\newenvironment{defi}{\refstepcounter{theorem} \medskip
\noindent{\it Definition \thetheorem:}}{\medskip}

\newcommand{\mar}[1]{}

\begin{document}
\hbox{}

{\parindent=0pt

{\large\bf Noether's second theorem for BRST symmetries}
\bigskip

{\sc D.Bashkirov}\footnote{Electronic mail: bashkir@phys.msu.ru}

{\sl Department of Theoretical Physics, Moscow State University,
117234 Moscow, Russia}

\medskip

{\sc G.Giachetta}\footnote{Electronic mail:
giovanni.giachetta@unicam.it}

{\sl Department of Mathematics and Informatics, University of
Camerino, 62032 Camerino (MC), Italy}

\medskip

{\sc L.Mangiarotti}\footnote{Electronic mail:
luigi.mangiarotti@unicam.it}

{\sl Department of Mathematics and Informatics, University of
Camerino, 62032 Camerino (MC), Italy}

\medskip

{\sc G. Sardanashvily}\footnote{Electronic mail:
sard@grav.phys.msu.su}

{\sl Department of Theoretical Physics, Moscow State University,
117234 Moscow, Russia}

\bigskip

We present Noether's second theorem for graded Lagrangian systems
of even and odd variables on an arbitrary body manifold $X$ in a
general case of BRST symmetries depending on derivatives of
dynamic variables and ghosts of any finite order. As a preliminary
step, Noether's second theorem for Lagrangian systems on fiber
bundles $Y\to X$ possessing gauge symmetries depending on
derivatives of dynamic variables and parameters of arbitrary order
is proved.

}

\bigskip
\bigskip

\noindent {\bf I. INTRODUCTION}
\bigskip

Different variants of Noether's second theorem state that, if a
Lagrangian admits symmetries depending on parameters, its
variational derivatives obey certain relations, called the Noether
identities. We present Noether's second theorem
 in the case of  BRST transformations
depending on derivatives (jets) of dynamic variables and ghosts of
arbitrary order. In particular, this is the case of the
field-antifield BRST theory and BV quantization.$^{1,2}$ A special
attention is paid to global aspects of Noether's second theorem as
a preliminary step of the global analysis of BV
quantization.$^{3,4}$

We start with a classical Lagrangian system on a fiber bundle
$Y\to X$ subject to even gauge transformations depending on even
dynamic variables, even parameters and their derivatives of any
order. For this purpose, we consider Lagrangian formalism on the
composite fiber bundle $E\to Y\to X$, where $E\to Y$ is a vector
bundle of gauge parameters. Accordingly, gauge transformations are
represented by a linear differential operator $\up$ on $E$ taking
its values in the vertical tangent bundle $VY$ of $Y\to X$
(Section II). The Noether identity for a Lagrangian $L$ is defined
as a differential operator $\Delta$ on the fiber bundle
(\ref{0548}) which takes its values in the density-dual
\mar{0630}\beq
E^*\op\ot_Y\op\w^n T^*X, \qquad n=\di X,\label{0630}
\eeq
of $E$ and whose kernel contains the image of the Euler--Lagrange
operator $\dl L$ of $L$, i.e., $\Delta\circ \dl L=0$ (Definition
\ref{0568}). Expressed in these terms, Noether's second theorem
(Section III, Theorem \ref{0510}) follows at once fom the
properties of differential operators on dual fiber bundles
(Appendix A, Theorem \ref{0610}). Namely, there exists the
intertwining operator $\eta(\up)=\Delta$, $\eta(\Delta)=\up$ such
that
\mar{0634,22}\ben
&& \eta(\eta(\up))=\up, \qquad \eta(\eta(\Delta))=\Delta,
\label{0634}\\
&& \eta(\up\circ\up')=\eta(\up')\circ\eta(\up), \qquad
\eta(\Delta'\circ\Delta)=\eta(\Delta)\circ\eta(\Delta').
\label{0622}
\een
The appropriate notions of reducible Noether identities and gauge
symmetries  are formulated, and their equivalence with respect to
the intertwining operator $\eta$ is proved (Section IV).

This formulation of Noether's second theorem is generalized to the
case of graded Lagrangian systems of even and odd variables and
BRST symmetries (Section VII, Theorem \ref{0716}). We describe odd
variables and their jets on an arbitrary smooth manifold $X$ as
generating elements of the structure ring of a graded manifold
whose body is $X$.$^{4,5}$ This definition differs from that of
jets of a graded fiber bundle,$^6$ but reproduces the heuristic
notion of jets of ghosts in the above mentioned field-antifield
BRST theory.$^{1,7}$

We consider BRST symmetries of a graded Lagrangian, i.e., its
nilpotent odd symmetries depending on ghosts as parameters
(Section VI). In particular, BRST symmetries come from the above
mentioned gauge symmetries by replacement of even parameters with
odd ghosts (Example \ref{0675}). In this case, the nilpotency
condition implies that original gauge symmetries form an algebra.

The key point is that, in order to define the Noether identity
associated to BRST symmetries, one should introduce antifields and
the Koszul--Tate differential. If a Noether identity is reducible,
$(0\leq k)$-stage ghosts and antighosts are called into play
(Section VIII), and we come to the complete tuple of fields,
ghosts and antifields in the field-antifield BRST theory.$^1$ We
however leave this theory outside the scope of the present work,
and keep an original graded Lagrangian independent of ghosts and
antifields.

\bigskip
\bigskip

\noindent {\bf II. GAUGE SYSTEMS ON FIBER BUNDLES}
\bigskip

Recall that an $r$-order Lagrangian on a fiber bundle $Y\to X$ is
defined as a
 density
\mar{0512}\beq
L=\cL\om:J^rY\to \op\w^nT^*X, \qquad \om=dx^1\w\cdots\w dx^n,
 \label{0512}
\eeq
on the $r$-order jet manifold $J^rY$ of sections of $Y\to X$. Jet
manifolds of $Y\to X$ make up the inverse system
\mar{5.10}\beq
X\op\longleftarrow^\pi Y\op\longleftarrow^{\pi^1_0} J^1Y
\longleftarrow \cdots J^{r-1}Y \op\longleftarrow^{\pi^r_{r-1}}
J^rY\longleftarrow\cdots. \label{5.10}
\eeq
In the sequel, the index $r=0$ stands for $Y$. Accordingly, we
have the direct system
\mar{5.7}\beq
\cO^*X\op\longrightarrow^{\pi^*} \cO^*Y
\op\longrightarrow^{\pi^1_0{}^*} \cO_1^*Y \ar\cdots \cO^*_{r-1}Y
\op\longrightarrow^{\pi^r_{r-1}{}^*}
 \cO_r^*Y \longrightarrow\cdots \label{5.7}
\eeq
of graded differential algebras (henceforth GDAs) $\cO_r^*Y$ of
exterior forms on jet manifolds $J^rY$ with respect to the
pull-back monomorphisms $\pi^r_{r-1}{}^*$. Its direct limit
 $\cO_\infty^*Y$ is a GDA
consisting of all exterior forms on finite order jet manifolds
modulo the pull-back identification.

The projective limit $(J^\infty Y, \pi^\infty_r:J^\infty Y\to
J^rY)$ of the inverse system (\ref{5.10}) is a Fr\'echet
manifold.$^8$ A bundle atlas $\{(U_Y;x^\la,y^i)\}$ of $Y\to X$
yields the coordinate atlas
\mar{jet1}\beq
\{((\pi^\infty_0)^{-1}(U_Y); x^\la, y^i_\La)\}, \qquad
{y'}^i_{\la+\La}=\frac{\dr x^\m}{\dr x'^\la}d_\m y'^i_\La, \qquad
0\leq|\La|, \label{jet1}
\eeq
of $J^\infty Y$, where $\La=(\la_k...\la_1)$ is a symmetric
multi-index, $\la+\La=(\la\la_k...\la_1)$,  and
\mar{5.177}\beq
d_\la = \dr_\la + \op\sum_{0\leq|\La|} y^i_{\la+\La}\dr_i^\La,
\qquad d_\La=d_{\la_r}\circ\cdots\circ d_{\la_1} \label{5.177}
\eeq
are the total derivatives. There is the restriction epimorphism
$\cO^*_\infty Y\to \cO^*_\infty U_Y$.  Therefore, $\cO^*_\infty Y$
can be written in a coordinate form where the horizontal one-forms
$\{dx^\la\}$ and the contact one-forms $\{\th^i_\La=dy^i_\La
-y^i_{\la+\La}dx^\la\}$ are generating elements of the
$\cO^0_\infty U_Y$-algebra $\cO^*_\infty U_Y$. Though $J^\infty Y$
is not a smooth manifold, elements of $\cO^*_\infty Y$ are
exterior forms on finite order jet manifolds and, therefore, their
coordinate transformations are smooth.

There is the canonical decomposition $\cO^*_\infty
Y=\oplus\cO^{k,m}_\infty Y$ of $\cO^*_\infty Y$ into $\cO^0_\infty
Y$-modules $\cO^{k,m}_\infty Y$ of $k$-contact and $m$-horizontal
forms together with the corresponding projectors $h_k:\cO^*_\infty
Y\to \cO^{k,*}_\infty Y$ and $h^m:\cO^*_\infty Y\to
\cO^{*,m}_\infty Y$. Accordingly, the exterior differential on
$\cO_\infty^* Y$ is split into the sum $d=d_H+d_V$ of the
nilpotent total and vertical differentials
\be
d_H(\f)= dx^\la\w d_\la\f, \qquad d_V(\f)=\th^i_\La \w
\dr^\La_i\f, \qquad \f\in\cO^*_\infty Y.
\ee
One also introduces the $\Bbb R$-module projector
\mar{r12}\beq
\vr=\op\sum_{0<k} \frac1k\ol\vr\circ h_k\circ h^n, \qquad
\ol\vr(\f)= \op\sum_{0\leq|\La|} (-1)^{\nm\La}\th^i\w
[d_\La(\dr^\La_i\rfloor\f)], \qquad \f\in \cO^{>0,n}_\infty Y,
\label{r12}
\eeq
of $\cO^*_\infty Y$ such that $\vr\circ d_H=0$, and the nilpotent
variational operator $\dl=\vr\circ d$ on $\cO^{*,n}_\infty Y$. Let
us put $\bE_k=\vr(\cO^{k,n}_\infty Y)$. Then the GDA $\cO^*_\infty
Y$ is split into the well-known variational bicomplex.$^{4,8-10}$
Here, we are concerned
 with its variational subcomplex
\mar{b317}\beq
0\to\Bbb R\to \cO^0_\infty Y \ar^{d_H}\cO^{0,1}_\infty Y\cdots
\op\ar^{d_H} \cO^{0,n}_\infty Y \op\ar^\dl \bE_1 \op\ar^\dl \bE_2
\ar \cdots\, \label{b317}
\eeq
and the subcomplex of one-contact forms
\mar{0623}\beq
 0 \to \cO^{1,0}_\infty Y\ar^{d_H} \cO^{1,1}_\infty Y
\cdots\ar^{d_H} \cO^{1,n}_\infty Y\ar^\vr \bE_1\to  0.
\label{0623}
\eeq
They possess the following cohomology.$^{3,4,11}$

\begin{theo} \label{g90} \mar{g90}
The cohomology of the variational complex (\ref{b317}) equals the
de Rham cohomology of $Y$.
\end{theo}

\begin{theo} \label{g90'} \mar{g90'}
The complex (\ref{0623}) is exact.
\end{theo}

Any finite order Lagrangian $L$ (\ref{0512}) is an element of
$\cO^{0,n}_\infty Y$, while
\mar{0513}\beq
\dl L=\cE_i\th^i\w\om=\op\sum_{0\leq|\La|}
(-1)^{|\La|}d_\La(\dr^\La_i \cL)\th^i\w\om\in \bE_1 \label{0513}
\eeq
is its Euler--Lagrange operator taking the values in the vector
bundle
\mar{0548}\beq
T^*Y\op\w_Y(\op\w^n T^*X)= V^*Y\op\ot_Y\op\w^n T^*X. \label{0548}
\eeq
The components $\cE_i$ of $\dl L$ are called the variational
derivatives. We further abbreviate $A\ap 0$ with an equality which
holds on-shell. This means that $A$ is an element of a module over
the ideal $I_L$ of the ring $\cO^0_\infty Y$ which is locally
generated by the variational derivatives $\cE_i$ (\ref{0513}) and
their total derivatives $d_\La\cE_i$. Thus, $I_L$ is a
differential ideal.

By virtue of Theorem \ref{g90}, every $\dl$-closed Lagrangian
$L\in\cO^{0,n}_\infty Y$ is the sum
\mar{t42}\beq
 L=h_0\psi + d_H\si,  \qquad \si\in
\cO^{0,n-1}_\infty Y, \label{t42}
\eeq
where $\psi$ is a closed $n$-form on $Y$. Theorem \ref{g90'}
provides the $\Bbb R$-module decomposition
\be
\cO^{1,n}_\infty Y=\bE_1\oplus d_H(\cO^{1,n-1}_\infty Y).
\ee
Given a Lagrangian $L\in \cO^{0,n}_\infty Y$, we have the
corresponding decomposition
\mar{+421}\beq
dL=\dl L-d_H\Xi \label{+421}
\eeq
where $\Xi_L=\Xi +L$ is a Lepagean equivalent of $L$.

Let $\gd\cO^0_\infty Y$ be the $\cO^0_\infty Y$-module of
derivations of the $\Bbb R$-ring $\cO^0_\infty Y$. Any $\vt\in
\gd\cO^0_\infty Y$ yields the graded derivation (the interior
product) $\vt\rfloor\f$ of the GDA $\cO^*_\infty Y$ given by the
relations
\be
&&\vt\rfloor df=\vt(f), \qquad  f\in \cO^0_\infty Y, \\
&& \vt\rfloor(\f\w\si)=(\vt\rfloor \f)\w\si
+(-1)^{|\f|}\f\w(\vt\rfloor\si), \qquad \f,\si\in \cO^*_\infty Y,
\ee
and its derivation (the Lie derivative)
\mar{0515}\ben
&& \bL_\vt\f=\vt\rfloor d\f+ d(\vt\rfloor\f), \qquad \f\in
\cO^*_\infty Y, \label{0515}\\
&& \bL_\vt(\f\w\f')=\bL_\vt(\f)\w\f' +\f\w\bL_\vt(\f'). \nonumber
\een

Relative to an atlas (\ref{jet1}), a derivation
$\vt\in\gd\cO^0_\infty$ reads$^4$
\mar{g3}\beq
\vt=\vt^\la \dr_\la + \vt^i\dr_i + \op\sum_{|\La|>0}\vt^i_\La
\dr^\La_i, \label{g3}
\eeq
where the tuple of derivations $\{\dr_\la,\dr^\La_i\}$ is defined
as the dual of the set $\{dx^\la, dy^i_\La\}$ of generating
elements for the $\cO^0_\infty$-algebra $\cO^*_\infty$ with
respect to the interior product $\rfloor$, and local functions
$\vt^\la, \vt^i, \vt^i_\La\in \cO^0_\infty$ obey the
transformation law
\mar{g71}\beq
\vt'^\la=\frac{\dr x'^\la}{\dr x^\m}\vt^\m,  \qquad
\vt'^i_\La=\op\sum_{|\Si|\leq|\La|}\frac{\dr y'^i_\La}{\dr
y^j_\Si}\vt^j_\Si + \frac{\dr y'^i_\La}{\dr x^\m}\up^\m.
\label{g71}
\eeq
Note that the tuple of derivations $\{\dr^\La_i\}$ is the dual of
the basis $\{\th^i_\La\}$ of contact forms.

A derivation $\vt$ is called contact if the Lie derivative
$\bL_\vt$ (\ref{0515}) preserves the contact ideal of the GDA
$\cO^*_\infty Y$ generated by contact forms. A derivation $\vt$
(\ref{g3}) is contact iff
\mar{g4}\beq
\vt^i_\La=d_\La(\vt^i-y^i_\m\vt^\m)+y^i_{\m+\La}\vt^\m, \qquad
0<|\La|. \label{g4}
\eeq
Any contact derivation admits the horizontal splitting
\mar{g5}\beq
\vt=\vt_H +\vt_V=\vt^\la d_\la + (\up^i\dr_i + \op\sum_{0<|\La|}
d_\La \up^i\dr_i^\La), \qquad \up^i= \vt^i-y^i_\m\vt^\m,
\label{g5}
\eeq
 relative to the canonical connection
$\nabla=dx^\la\ot d_\la$ on the $C^\infty(X)$-ring
$\cO^0_\infty$.$^{5,12}$ Its vertical part $\vt_V$  is completely
determined by the first summand
\mar{0641}\beq
\up=\up^i(x^\la,y^i_\La)\dr_i, \qquad 0\leq |\La|\leq k.
\label{0641}
\eeq
This is a section of the pull-back $VY\op\times_Y J^kY\to J^kY$ of
the vertical tangent bundle $VY\to Y$ onto $J^kY$,$^{13}$ i.e.,
$\up$ (\ref{0641}) is a $k$-order $VY$-valued differential
operator on $Y$ (see Appendix A). One calls this differential
operator the {\it generalized vector field} on $Y$.

\begin{prop}  \label{g75} \mar{g75}
It follows from the splitting (\ref{+421}) that the Lie derivative
of a Lagrangian $L$ (\ref{0512}) along a contact derivation $\vt$
(\ref{g5}) fulfills the first variational formula
\mar{g8'}\beq
\bL_\vt L= \up\rfloor\dl L +d_H(h_0(\vt\rfloor\Xi_L)) +\cL d_V
(\vt_H\rfloor\om), \label{g8'}
\eeq
where $\Xi_L$ is a Lepagean equivalent of $L$.$^4$
\end{prop}

A contact derivation $\vt$ (\ref{g5}) is called  variational if
the Lie derivative (\ref{g8'}) is $d_H$-exact, i.e., $\bL_\vt
L=d_H\si$, $\si\in \cO^{0,n-1}_\infty$. A glance at the expression
(\ref{g8'}) shows that: (i) a contact derivation $\vt$ is
variational only if it is projected onto $X$ (i.e., its components
$\vt^\la$ depend only on coordinates on $X$), (ii) $\vt$ is
variational iff its vertical part $\vt_V$ is variational, (iii) it
is variational if $\up\rfloor \dl L$ is $d_H$-exact.

By virtue of item (ii), we can restrict our consideration to
vertical contact derivations $\vt=\vt_V$. A generalized vector
field $\up$ (\ref{0641}) is called a {\it variational symmetry} of
a Lagrangian $L$ if it generates a variational vertical contact
derivation.

One can also consider locally-variational contact derivations when
the Lie derivative (\ref{g8'}) is $\dl$-closed, but any
locally-variational gauge symmetry is always variational (see
Remark \ref{0700} below).

Turn now to the notion of a gauge symmetry. A Lagrangian system on
a fiber bundle $Y\to X$ is said to be a gauge theory if its
Lagrangian $L$ admits a family of variational symmetries
parameterized by elements of a vector bundle $E\to Y$ as follows.

Let $E\to Y$ be a vector bundle coordinated by
$(x^\la,y^i,\xi^r)$. Given a Lagrangian $L$ on $Y$, let us
consider its pull-back, say again $L$, onto $E$. Let $\vt_E$ be a
vertical contact derivation of the $\Bbb R$-ring $\cO^0_\infty E$
whose restriction
\mar{0508}\beq
\vt=\vt_E|_{\cO^0_\infty Y}=
\op\sum_{0\leq|\La|}d_\La\up^i\dr_i^\La \label{0508}
\eeq
to $\cO^0_\infty Y\subset \cO^0_\infty E$ is linear in coordinates
$\xi^r_\Xi$. It is determined by a generalized vector field (i.e.,
a $VE$-valued differential operator) $\up_E$ on $E$ whose
canonical projection
\be
\up:J^kE\ar^{\up_E} VE\to E\op\times_Y VY
\ee
(see the exact sequence (\ref{0640}) below) is a linear
$VY$-valued differential operator
\mar{0509}\beq
\up= \op\sum_{0\leq|\Xi|\leq
m}\up^{i,\Xi}_r(x^\la,y^i_\Si)\xi^r_\Xi \dr_i \label{0509}
\eeq
on $E$. Let $\vt_E$ be a variational symmetry of a Lagrangian $L$
on $E$, i.e.,
\mar{0552}\beq
\up_E\rfloor \dl L=\up\rfloor \dl L=d_H\si. \label{0552}
\eeq
Then one says that $\up$ (\ref{0509}) is a {\it gauge symmetry} of
a Lagrangian $L$ on $Y$.

\begin{rem} \label{0700} \mar{0700}
As was mentioned above, any locally-variational gauge symmetry
$\vt_E$, when the Lie derivative $\bL_{\vt_E} L$ is $\dl$-closed,
is variational. By virtue of Theorem \ref{g90}, $\bL_{\vt_E} L$
takes the form (\ref{t42}) where $\psi$ is a closed form on $E$.
Since $E\to Y$ is a vector bundle, $Y$ is a strong deformation
retract of $E$ and, consequently, the de Rham cohomology of $E$
equals that of $Y$. Then any closed form on $E$ is the sum of the
pull-back of a closed form on $Y$ and an exact form on $E$. The
former is independent of fiber coordinates $\xi^r$ on $E\to Y$.
Since the Lie derivative $\bL_\vt L$ is linear in $\xi^r_\La$, it
is always $d_H$-exact, i.e., $\vt_E$ is variational.
\end{rem}

\bigskip
\bigskip

\noindent {\bf III. NOETHER'S SECOND THEOREM I}
\bigskip

Let us start with the notion of the Noether identity.

\begin{defi} \label{0568} \mar{0568}
Given a Lagrangian $L$ (\ref{0512}) and its Euler--Lagrange
operator $\dl L$ (\ref{0513}), let $E\to Y$ be a vector bundle and
$\Delta$ a linear differential operator of order $0\leq m$ on the
vector bundle (\ref{0548}) with the values in the density-dual
$\ol E^*$ (\ref{0630}) of $E$ such that
\mar{0549}\beq
\Delta\circ \dl L=0. \label{0549}
\eeq
This condition is called the {\it Noether identity}, and $\Delta$
is the {\it Noether operator}.
\end{defi}

Given bundle coordinates $(x^\la,y^i, \ol y_i)$ on the fiber
bundle (\ref{0548}) and  $(x^\la,y^i,\xi^r)$ on $E$, a Noether
operator $\Delta$ in Definition \ref{0568} is represented by the
 density
\mar{0547}\beq
\Delta= \Delta_r\xi^r\om= \op\sum_{0\leq|\La|\leq m}
\Delta^{i,\La}_r(x^\la,y^j_\Si)\ol y_{\La i} \xi^r\om\in
\cO^{0,n}_\infty[E\op\times_Y V^*Y], \qquad 0\leq|\Si|\leq m,
\label{0547}
\eeq
(see Appendix A). Then the Noether identity (\ref{0549}) takes the
coordinate form
\mar{0550}\beq
[\op\sum_{0\leq|\La|\leq m} \Delta^{i,\La}_r d_\La \cE_i]
\xi^r\om=0, \label{0550}
\eeq
where $\cE_i$ are variational derivatives (\ref{0513}).

\begin{rem} We further use the relations
\mar{0606a-d}\ben
&& \op\sum_{0\leq |\La|\leq k}B^\La d_\La A'= \op\sum_{0\leq
|\La|\leq k} (-1)^{|\La|}d_\La (B^\La) A' +
d_H\si, \label{0606a}\\
&& \op\sum_{0\leq |\La|\leq k} (-1)^{|\La|}d_\La(B^\La A)=
\op\sum_{0\leq |\La|\leq k} \eta (B)^\La d_\La A, \label{0606b}
\\ && \eta (B)^\La = \op\sum_{0\leq|\Si|\leq
k-|\La|}(-1)^{|\Si+\La|} C^{|\Si|}_{|\Si+\La|} d_\Si B^{\Si+\La},
\qquad C^a_b=\frac{b!}{a!(b-a)!}, \label{0606c}\\
&& (\eta\circ\eta)(B)^\La=B^\La, \label{0606d}
\een
for any exterior forms $A'\in\cO^{*,n}_\infty Z$,  $A\in\cO^*_\infty Z$
and any local function
$B^\La\in \cO^0_\infty Z$ on jet manifolds of a fiber bundle $Z\to
X$.
\end{rem}

\begin{theo} \label{0510} \mar{0510}
If a Lagrangian $L$ (\ref{0512}) admits a gauge symmetry $\up$
(\ref{0509}), its Euler--Lagrange operator obeys the Noether
identity (\ref{0550}) where
\mar{0511}\beq
\Delta^{i,\La}_r =\eta(\up)^{i,\La}_r=\op\sum_{0\leq|\Si|\leq
m-|\La|}(-1)^{|\Si+\La|}C^{|\Si|}_{|\Si+\La|} d_\Si
\up^{i,\Si+\La}_r. \label{0511}
\eeq
Conversely, if the Euler--Lagrange operator of a Lagrangian $L$
obeys the Noether identity (\ref{0550}), this Lagrangian admits a
gauge symmetry  $\up$ (\ref{0509}) where
\mar{0638}\beq
\up^{i,\La}_r =\eta(\Delta)^{i,\La}_r=\op\sum_{0\leq|\Si|\leq
m-|\La|}(-1)^{|\Si+\La|} C^{|\Si|}_{|\Si+\La|}d_\Si
\Delta^{i,\Si+\La}_r. \label{0638}
\eeq
The relations (\ref{0634}) hold.
\end{theo}

\begin{proof} Given a differential operator $\up$ (\ref{0509}), the
operator $\Delta=\eta(\up)$ expressed in the coordinate form
(\ref{0511}) is defined in accordance with Theorem \ref{0610}.
Since the  density
\be
\up\rfloor\dl L=\up^i\cE_i\om=\op\sum_{0\leq|\Xi|\leq m}
\up^{i,\Xi}_r\xi^r_\Xi\cE_i\om
\ee
is $d_H$-exact, the Noether identity
\be
\dl(\up\rfloor\dl L)=\eta(\up)\circ \dl L=0
\ee
holds. Conversely, any Noether operator $\Delta$ (\ref{0547})
defines the $VY$-valued differential operator $\up=\eta(\Delta)$
on $E$ expressed in the coordinate form (\ref{0638}). This
differential operator gives rise to a $VE$-valued differential
operator (i.e., a generalized vector field) $\up_E$ on $E$ and,
thus, defines a contact derivation $\vt_E$ of $\cO^0_\infty E$.
Indeed, let us consider the exact sequence of vector bundles
\mar{0640}\beq
0\to V_YE\to VE\to E\op\times_Y VY\to 0, \label{0640}
\eeq
where $V_YE$ is the vertical tangent bundle of  $E\to Y$. Any
splitting $\G$ of this exact sequence lifts $\up$ to the
generalized vector field $\up_E=\G\circ\up$ on $E$, but the Lie
derivative $\bL_{\vt_E}L$ is independent of the choice of a
splitting $\G$. Due to the Noether identity (\ref{0550}), we
obtain
\be
&&0=\op\sum_{0\leq|\La|\leq m}
\xi^r\Delta^{i,\La}_rd_\La\cE_i\om=\op\sum_{0\leq|\La|\leq m}
(-1)^{|\La|}d_\La(\xi^r \Delta^{i,\La}_r)\cE_i\om +
d_H\si=\\
&& \qquad \op\sum_{0\leq|\Xi|\leq m}\up^{i,\Xi}_r\xi^r_\Xi\cE_i\om
+d_H\si=\up\rfloor \dl L+d_H\si,
\ee
i.e., $\up$ is a gauge symmetry of $L$. Due to the equality
(\ref{0606d}), the relations (\ref{0634}) hold.
\end{proof}

\begin{ex} \label{0655} \mar{0655} If a gauge symmetry
\mar{0656}\beq
\up=(\up_r^i\xi^r +\up^{i,\m}_r\xi^r_\m)\dr_i \label{0656}
\eeq
is of first jet order in parameters,  the corresponding Noether
operator and Noether identity read
\mar{0657,8}\ben
&& \Delta^i_r=\up^i_r -d_\m \up^{i,\m}_r,\qquad \Delta^{i,\m}_r=-
\up^{i,\m}_r,\label{0657}\\
&& [\up^i_r\cE_i - d_\m(\up^{i,\m}_r\cE_i)]\xi^r\om=0.
\label{0658}
\een
\end{ex}

Any Lagrangian $L$ has gauge symmetries. In particular, there
always exist trivial gauge symmetries
\be
\up=\op\sum_\La \eta(M)^{i,\La}_r\xi^r_\La, \qquad
M^{i,\La}_r=\op\sum_\Si T^{i,j,\La,\Si}d_\Si\cE_j,  \qquad
T_r^{j,i,\La,\Si}=-T_r^{i,j,\Si,\La},
\ee
corresponding to the trivial Noether identity
\be
\op\sum_{\Si,\La}T_r^{j,i,\La,\Si}d_\Si\cE_j d_\La\cE_i=0.
\ee
Furthermore, given a gauge symmetry $\up$ (\ref{0509}), let $h$ be
a linear differential operator on some vector bundle $E'\to Y$,
coordinated by $(x^\la,y^i,\xi'^s)$, with values in the vector
bundle $E$. Then the composition
\be
\up'_0=\up\circ h=\up'^{i,\La}_s\xi'^s_\La\dr_i, \qquad
\up'^{i,\La}_s=\op\sum_{\Xi+\Xi'=\La}\op\sum_{0\leq|\Si|\leq
m-|\Xi|} \up^{i,\Xi+\Si}_rd_\Si h^{r,\Xi'}_s,
\ee
is a variational symmetry of the pull-back onto $E'$ of a
Lagrangian $L$ on $Y$, i.e., a gauge symmetry of $L$. In view of
this ambiguity, we agree to say that a gauge symmetry $\up$
(\ref{0509}) of a Lagrangian $L$ is complete if a different gauge
symmetry $\up'_0$ of $L$ factors through $\up$ as
\be
\up'_0=\up\circ h + T, \qquad T\ap 0.
\ee
A complete gauge symmetry always exists, but the vector bundle of
its parameters need not be finite-dimensional.

Accordingly, given  the Noether operator (\ref{0547}), let $H$ be
a linear differential operator on $\ol E^*$ with values in the
density-dual $\ol E'^*$ (\ref{0630}) of some vector bundle $E'\to
Y$. Then the composition $\Delta'=H\circ\Delta$ ia also a Noether
operator. We agree to call the Noether operator (\ref{0547})
complete if a different Noether operator $\Delta'$ factors through
$\Delta$ as
\be
\Delta'=H\circ\Delta + F, \qquad F\ap 0.
\ee

\begin{prop} \label{0646} \mar{0646}
A gauge symmetry $\up$ of a Lagrangian $L$ is complete iff so is
the associated Noether operator.
\end{prop}

\begin{proof} The proof follows at once from Proposition
\ref{0621} in Appendix A. Given a gauge symmetry $\up$ of $L$, let
$\up'_0$ be a different gauge symmetry. If $\eta(\up)$ is a
complete Noether operator, then
\be
\eta(\up'_0)=H\circ\eta(\up) + F, \qquad F\ap 0,
\ee
and, by virtue of the relations (\ref{0622}), we have
\be
\up'_0=\up\circ\eta(H) +\eta(F),
\ee
where $\eta(F)\ap 0$ because $I_L$ is a differential ideal. The
converse is similarly proved.
\end{proof}

\begin{ex} \label{0650} \mar{0650}
Let us consider the gauge theory of principal connections on a
principal bundle $P\to X$ with a structure Lie group $G$.$^{12}$
These connections are represented by sections of the quotient
\mar{0654}\beq
C=J^1P/G\to X,\label{0654}
\eeq
called the bundle of principal connections. This is an affine
bundle coordinated by $(x^\la, a^r_\la)$ such that, given a
section $A$ of $C\to X$, its components $A^r_\la=a^r_\la\circ A$
are coefficients of the familiar local connection form (i.e.,
gauge potentials). Let $J^\infty C$ be the infinite order jet
manifold of $C\to X$ coordinated by $(x^\la,a^r_{\La\la})$, $0\leq
|\La|$. We consider the GDA $\cO^*_\infty C$. Infinitesimal
generators of local one-parameter groups of automorphisms of a
principal bundle $P$ are $G$-invariant projectable vector fields
on $P\to X$. They are associated to sections of the vector bundle
$T_GP=TP/G\to X$. This bundle is endowed with the coordinates
$(x^\la,\tau^\la=\dot x^\la,\xi^r)$ with respect to the fiber
bases $\{\dr_\la, e_r\}$ for $T_GP$, where $\{e_r\}$ is the basis
for the right Lie algebra $\ccG$ of $G$ such that
$[e_p,e_q]=c^r_{pq}e_r.$ If
\mar{0652}\beq
 u=u^\la\dr_\la
+u^r e_r, \qquad v=v^\la\dr_\la +v^r e_r, \label{0652}
\eeq
are sections of $T_GP\to X$, their bracket reads
\mar{0651}\beq
[u,v]=(u^\m\dr_\m v^\la -v^\m\dr_\m u^\la)\dr_\la +(u^\la\dr_\la
v^r - v^\la\dr_\la u^r +c^r_{pq}u^pv^q)e_r. \label{0651}
\eeq
Any section $u$ of the vector bundle $T_GP\to X$ yields the vector
field
\mar{0653}\beq
u_C=u^\la\dr_\la +(c^r_{pq}a^p_\la u^q +\dr_\la u^r -a^r_\m\dr_\la
u^\m)\dr^\la_r \label{0653}
\eeq
on the bundle of principal connections $C$ (\ref{0654}). It is an
infinitesimal generator of a one-parameter group of automorphisms
of $C$.$^{12}$ Let us consider the bundle product
\mar{0659}\beq
E=C\op\times_X T_GP, \label{0659}
\eeq
coordinated by $(x^\la, \tau^\la, \xi^r, a^r_\la)$. It can be
provided with the generalized vector field
\mar{0660}\beq
\up_E= \up=(c^r_{pq}a^p_\la \xi^q + \xi^r_\la
-a^r_\m\tau^\m_\la-\tau^\m a_{\m\la}^r)\dr^\la_r. \label{0660}
\eeq
For instance, this is a gauge symmetry of the global Chern--Simons
Lagrangian.$^{14}$  Let us consider a subbundle $V_GP=VP/G\to X$
of the vector bundle $T_GX$ coordinated by $(x^\la, \xi^r)$. Its
sections $u=u^re_r$ are infinitesimal generators of vertical
automorphisms of $P$. Let us restrict the bundle product
(\ref{0659}) to
\mar{0661}\beq
E=C\op\times_X V_GP. \label{0661}
\eeq
It is provided with the generalized vector field
\mar{0662}\beq
\up_E= \up=(c^r_{pq}a^p_\la \xi^q + \xi^r_\la)\dr^\la_r.
\label{0662}
\eeq
This is a gauge symmetry of the Yang--Mills Lagrangians,$^{14}$
and yields the well-known Noether identity
\be
[c^r_{pq}a^p_\la\cE_r^\la - d_\la(\cE_q^\la)]\xi^q\om=0.
\ee
\end{ex}

\bigskip
\bigskip

\noindent {\bf IV. REDUCIBLE GAUGE THEORIES}
\bigskip

Recall that the notion of a reducible Noether identity has come
from that of a reducible constraint,$^{15}$ but it involves
differential relations.

\begin{defi} \label{0573} \mar{0573} A complete Noether operator
$\Delta\not\ap 0$ (\ref{0547}) and the corresponding Noether
identity (\ref{0549}) are said to be {\it $N$-stage reducible}
($N=0,1,\ldots$) if there exist vector bundles $E_k\to Y$ and
differential operators $\Delta_k$, $k=0,\ldots,N$, such that:

(i)  $\Delta_k$ is a linear differential operator on the
density-dual $\ol E^*_{k-1}$ of $E_{k-1}$ with values in the
density-dual $\ol E^*_k$ of $E_k$, where $E_{-1}=E$;

(ii) $\Delta_k\not\ap 0$ for all $k=0,\ldots,N$;

(iii) $\Delta_k\circ \Delta_{k-1}\ap 0$ for all $k=0,\ldots,N$,
where $\Delta_{-1}=\Delta$;

(iv) if $\Delta'_k$ is another differential operator possessing
these properties, then it factors through $\Delta_k$ on-shell.
\end{defi}

In particular, a zero-stage reducible Noether operator is called
reducible. In this case, given bundle coordinates
$(x^\la,y^i,\ol\xi_r)$ on $\ol E^*$ and $(x^\la,y^i,\xi^{r_0})$ on
$E_0$, a differential operator $\Delta_0$ reads
\mar{0576}\beq
\Delta_0=\op\sum_{0\leq|\Xi|\leq m_0} \Delta^{r,\Xi}_{r_0}
\ol\xi_{\Xi r}\xi^{r_0}\om. \label{0576}
\eeq
Then the reduction condition $\Delta_0\circ\Delta\ap 0$ takes the
coordinate form
\mar{0570}\beq
\op\sum_{0\leq|\Xi|\leq m_0}
\Delta^{r,\Xi}_{r_0}d_\Xi(\op\sum_{0\leq|\La|\leq
m}\Delta^{i,\La}_r\ol y_{\La i})\xi^{r_0}\om\ap 0, \label{0570}
\eeq
i.e., the left hand-side of this expression takes the form
\be
\op\sum_{0\leq|\Si|\leq m_0+m}M^{i,\Si}_{r_0} \ol y_{\Si
i}\xi^{r_0}\om,
\ee
where all the coefficients $M^{i,\Si}_{r_0}$ belong to the ideal
$I_L$.

\begin{defi} \label{0574} \mar{0574} A complete gauge
symmetry $\up\not\ap 0$ (\ref{0509}) is said to be $N$-stage
reducible if  there exist vector bundles $E_k$ and differential
operators $\up^k$, $k=0,\ldots,N$, such that:

(i)  $\up^k$ is a linear differential operator on the vector
bundle $E_k$ with values in the vector bundle $E_{k-1}$;

(ii) $\up^k\not\ap 0$ for all $k=0,\ldots,N$;

(iii) $\up^{k-1}\circ \up^k\ap 0$ for all $k=0,\ldots,N$, where
$\up^k$, $k=-1$, stands for $\up$;

(iv) if $\up'^k$ is another differential operator possessing these
properties, then $\up'^k$ factors through $\up^k$ on-shell.
\end{defi}

\begin{theo} \label{0580} \mar{0580}
A gauge symmetry $\up$ is $N$-stage reducible iff so is the
associated Noether identity.
\end{theo}

\begin{proof}
The proof follows at once from Theorem \ref{0610} and Proposition
\ref{0621}. Let us put $\Delta_k=\eta(\up^k)$, $k=0,\ldots,N$. If
$\up^k\ap 0$, then $\eta(\up^k)\ap 0$ because $I_L$ is a
differential ideal. By the same reason, if $\up^{k-1}$ and $\up^k$
obey the reduction condition $\up^{k-1}\circ \up^k\ap 0$, then
\be
\eta(\up^{k-1}\circ \up^k)=\eta(\up^k)\circ\eta(\up^{k-1})\ap 0.
\ee
The converse  is justified in the same way. The equivalence of the
conditions in items (iv) of Definitions \ref{0573} and \ref{0574}
is proved similarly to that in Proposition \ref{0646}.
\end{proof}

\bigskip
\bigskip

\noindent {\bf V. GRADED LAGRANGIAN SYSTEMS}
\bigskip

Recall that, by virtue of Batchelor's theorem,$^{16}$ any graded
manifold $(X,\gA)$ with a body $X$ is isomorphic to the one whose
structure sheaf $\gA_Q$ is formed by germs of sections of the
exterior product
\mar{g80}\beq
\w Q^*=\Bbb R\op\oplus_X Q^*\op\oplus_X\op\w^2
Q^*\op\oplus_X\cdots, \label{g80}
\eeq
where $Q^*$ is the dual of some real vector bundle $Q\to X$ of
fiber dimension $m$. In field models, a vector bundle $Q$ is
usually given from the beginning. Therefore, we consider graded
manifolds $(X,\gA_Q)$ where Batchelor's isomorphism is fixed, and
call $(X,\gA_Q)$ the simple graded manifold constructed from $Q$.
The structure ring $\cA_Q$ of sections of $\gA_Q$ consists of
sections of the exterior bundle (\ref{g80}) called graded
functions. Given bundle coordinates $(x^\la,q^a)$ on $Q$ with
transition functions $q'^a=\rho^a_b q^b$, let $\{c^a\}$ be the
corresponding fiber bases for $Q^*\to X$, together with transition
functions $c'^a=\rho^a_bc^b$. Then $(x^\la, c^a)$ is called the
local basis for the graded manifold $(X,\gA_Q)$. With respect to
this basis, graded functions read
\be
f=\op\sum_{k=0}^m \frac1{k!}f_{a_1\ldots a_k}c^{a_1}\cdots
c^{a_k},
\ee
where $f_{a_1\cdots a_k}$ are local smooth real functions on $X$,
and we omit the symbol of the exterior product of elements $c^a$.

Given a graded manifold $(X,\gA_Q)$, let $\gd\cA_Q$ be the
$\cA_Q$-module of $\Bbb Z_2$-graded derivations of the $\Bbb
Z_2$-graded ring $\cA_Q$, i.e.,
\be
u(ff')=u(f)f'+(-1)^{[u][f]}fu (f'), \qquad u\in\gd\cA_Q, \qquad
f,f'\in \cA_Q,
\ee
where $[.]$ denotes the Grassmann parity. Its elements are called
$\Bbb Z_2$-graded (or, simply, graded) vector fields on
$(X,\gA_Q)$. Due to the canonical splitting $VQ= Q\times Q$, the
vertical tangent bundle $VQ\to Q$ of $Q\to X$ can be provided with
the fiber basis $\{\dr_a\}$ which is the dual of $\{c^a\}$. Then a
graded vector field takes the local form $u= u^\la\dr_\la +
u^a\dr_a$, where $u^\la, u^a$ are local graded functions. It acts
on $\cA_Q$ by the rule
\mar{cmp50'}\beq
u(f_{a\ldots b}c^a\cdots c^b)=u^\la\dr_\la(f_{a\ldots b})c^a\cdots
c^b +u^d f_{a\ldots b}\dr_d\rfloor (c^a\cdots c^b). \label{cmp50'}
\eeq
This rule implies the corresponding transformation law
\be
u'^\la =u^\la, \qquad u'^a=\rho^a_ju^j +
u^\la\dr_\la(\rho^a_j)c^j.
\ee
Then one can show$^{5,12}$ that graded vector fields on a simple
graded manifold can be represented by sections of the vector
bundle $\cV_Q\to X$ which is locally isomorphic to the vector
bundle $\w Q^*\ot_X(Q\oplus_X TX)$.

Using this fact, we can introduce graded exterior forms on the
simple graded manifold $(X,\gA_Q)$ as sections of the exterior
bundle $\op\w\cV^*_Q$, where $\cV^*_Q\to  X$ is the $\w Q^*$-dual
of $\cV_Q$. They are characterized both by the Grassmann parity
and the familiar form degree. Relative to the dual local bases
$\{dx^\la\}$ for $T^*X$ and $\{dc^b\}$ for $Q^*$, graded one-forms
read
\be
\f=\f_\la dx^\la + \f_adc^a,\qquad \f'_a=\rho^{-1}{}_a^b\f_b,
\qquad \f'_\la=\f_\la +\rho^{-1}{}_a^b\dr_\la(\rho^a_j)\f_bc^j,
\ee
where $dx^\la$ are even and $dc^b$ are odd. The duality morphism
is given by the interior product
\be
u\rfloor \f=u^\la\f_\la + (-1)^{[\f_a]}u^a\f_a.
\ee
Graded exterior forms constitute the bigraded differential algebra
(henceforth BGDA) $\cC^*_Q$ with respect to the bigraded exterior
product
\be
\f\w\f' =(-1)^{|\f||\f'| +[\f][\f']}\f'\w \f, \qquad \f,\f'\in
\cC^*_Q,
\ee
and the exterior differential
\be
 d(\f\w\f')= d\f\w\f' +(-1)^{|\f|}\f\w d\f', \qquad
\f,\f'\in \cC^*_Q.
\ee

Since the jet bundle $J^rQ\to X$ of a vector bundle $Q\to X$ is a
vector bundle, let us consider the simple graded manifold
$(X,\gA_{J^rQ})$ constructed from $J^rQ\to X$. Its local basis is
$\{x^\la,c^a_\La\}$, $0\leq |\La|\leq r$, together with the
transition functions
\mar{+471}\beq
c'^a_{\la +\La}=d_\la(\rho^a_j c^j_\La), \qquad d_\la=\dr_\la +
\op\sum_{|\La|<r}c^a_{\la+\La} \dr_a^\La, \label{+471}
\eeq
where the graded derivations $\dr_a^\La$ are the duals of
$c^a_\La$. Let $\cC^*_{J^rQ}$ be the BGDA of graded exterior forms
on the graded manifold $(X,\gA_{J^rQ})$. A linear bundle morphism
$\pi^r_{r-1}:J^rQ \to J^{r-1}Q$ yields the corresponding
monomorphism of BGDAs $\cC^*_{J^{r-1}Q}\to \cC^*_{J^rQ}$. Hence,
there is the direct system of BGDAs
\mar{g205}\beq
\cC^*_Q\ar^{\pi^{1*}_0} \cC^*_{J^1Q}\cdots
\ar^{\pi^r_{r-1}{}^*}\cC^*_{J^rQ}\ar\cdots. \label{g205}
\eeq
Its direct limit $\cC^*_\infty Q$ consists of graded exterior
forms on graded manifolds $(X,\gA_{J^rQ})$, $r\in\Bbb N$, modulo
the pull-back identification, and it inherits the BGDA operations
intertwined by the monomorphisms $\pi^r_{r-1}{}^*$. It is locally
a free $C^\infty(X)$-algebra locally generated by the elements
$(1, c^a_\La, dx^\la,\th^a_\La=dc^a_\La -c^a_{\la +\La}dx^\la)$,
$0\leq|\La|$, where $c^a_\La$ and $\th^a_\La$ are odd.

 In order to regard even and odd dynamic variables  on the
same footing,  let $Y\to X$ be hereafter an affine bundle, and let
$\cP^*_\infty Y\subset \cO^*_\infty Y$ be the
$C^\infty(X)$-subalgebra of exterior forms whose coefficients are
polynomial in the fiber coordinates on jet bundles $J^r Y\to X$.
This notion is intrinsic since any element of $\cO^*_\infty Y$ is
an exterior form on some finite order jet manifold and all jet
bundles $J^r Y\to X$ are affine. One can think of the GDA
$\cP^*_\infty Y$ as being the BGDA whose elements are even. Let us
consider the product
\mar{0670}\beq
\cS^*_\infty[Q;Y]=\cC_\infty^*Q\w\cP^*_\infty Y \label{0670}
\eeq
of bigraded algebras $\cC_\infty^*Q$ and $\cP^*_\infty Y$ over
their common graded subalgebra $\cO^*X$ of exterior forms on
$X$.$^4$ It consists of the elements
\be
\op\sum_i \psi_i\ot\f_i, \qquad \op\sum_i \f_i\ot\psi_i, \qquad
\psi\in \cC^*_\infty Q, \qquad \f\in \cP^*_\infty Y,
\ee
modulo the commutation relations
\mar{0442}\ben
&&\psi\ot\f=(-1)^{|\psi||\f|}\f\ot\psi, \qquad
\psi\in \cC^*_\infty Q, \qquad \f\in \cP^*_\infty Y, \label{0442}\\
&& (\psi\w\si)\ot\f=\psi\ot(\si\w\f), \qquad \si\in \cO^*X.
\nonumber
\een
These elements are  endowed with the total form degree $|\psi\ot
\f|=|\psi|+|\f|$ and the total Grassmann parity
$[\psi\ot\f]=[\psi]$. Their multiplication
\mar{0440}\beq
(\psi\ot\f)\w(\psi'\ot\f'):=(-1)^{|\psi'||\f|}(\psi\w\psi')\ot
(\f\w\f'). \label{0440}
\eeq
obeys the relation
\be
\vf\w\vf' =(-1)^{|\vf||\vf'| +[\vf][\vf']}\vf'\w \vf, \qquad
\vf,\vf'\in \cS^*_\infty[Q;Y],
\ee
and makes $\cS^*_\infty[Q;Y]$ (\ref{0670}) into a bigraded
$C^\infty (X)$-algebra.  For instance, elements of the ring
$\cS^0_\infty[Q;Y]$ are polynomials of odd $c^a_\La$ and even
$y^i_\La$ with coefficients in $C^\infty(X)$.

The algebra $\cS^*_\infty[Q;Y]$ is provided with the exterior
differential
\mar{0441}\beq
d(\psi\ot\f):=(d_\cC\psi)\ot\f +(-1)^{|\psi|}\psi\ot(d_\cP\f),
\qquad \psi\in \cC^*_\infty, \qquad \f\in \cP^*_\infty,
\label{0441}
\eeq
where $d_\cC$ and $d_\cP$ are exterior differentials on the
differential algebras $\cC^*_\infty Q$ and $\cP^*_\infty Y$,
respectively. It obeys the relations
\be
 d(\vf\w\vf')= d\vf\w\vf' +(-1)^{|\vf|}\vf\w d\vf', \qquad
\vf,\vf'\in \cS^*_\infty[Q;Y],
\ee
and makes $\cS^*_\infty[Q;Y]$ into a BGDA, which is locally
generated by the elements
\be
(1, c^a_\La, y^i_\La,
dx^\la,\th^a_\La=dc^a_\La-c^a_{\la+\La}dx^\la,\th^i_\La=
dy^i_\La-y^i_{\la+\La}dx^\la), \qquad 0\leq |\La|,
\ee
where $c^a_\La$, $\th^a_\La$ are odd and $y^i_\La$, $dx^\la$,
$\th^i_\La$ are even. The cohomology of its de Rham complex
\mar{g110}\beq
0\to\Bbb R\ar \cS^0_\infty[Q;Y]\ar^d \cS^1_\infty[Q;Y]\cdots
\ar^d\cS^k_\infty[Q;Y] \ar\cdots \label{g110}
\eeq
equals the de Rham cohomology $H^*(X)$ of $X$.$^4$ We agree to
call elements of $\cS^*_\infty[Q;Y]$ the {\it graded exterior
forms} on $X$.

Hereafter, let the collective symbols $s^A_\La$ and $\th^A_\La$
stand both for even and odd generating elements $c^a_\La$,
$y^i_\La$, $\th^a_\La$, $\th^i_\La$ of the $C^\infty(X)$-algebra
$\cS^*_\infty[Q;Y]$ which, thus, is locally generated by
$(1,s^A_\La, dx^\la, \th^A_\La)$, $|\La|\geq 0$. Since
$s^A_\La=d_\La s^A$ and $\th^A_\La=ds^A_\La+ s^A_{\la+\La}dx^\la$,
the BGDA $\cS^*_\infty[Q;Y]$ is completely specified by the
elements $s^A$ together with their transition functions.
Therefore, we agree to call $\{s^A\}$ the {\it local basis} for
$\cS^*_\infty[Q;Y]$.

Similarly to $\cO^*_\infty Y$, the BGDA $\cS^*_\infty[Q;Y]$ is
decomposed into $\cS^0_\infty[Q;Y]$-modules
$\cS^{k,r}_\infty[Q;Y]$ of $k$-contact and $r$-horizontal graded
forms together with the corresponding projections $h_k$ and $h^r$.
Accordingly, the exterior differential $d$ (\ref{0441}) on
$\cS^*_\infty[Q;Y]$ is split into the sum $d=d_H+d_V$ of the total
and vertical differentials
\be
d_H(\f)=dx^\la\w d_\la(\f), \qquad d_V(\f)=\th^A_\La\w\dr^\La_A
\f, \qquad \f\in \cS^*_\infty[Q;Y].
\ee
The projection endomorphism $\vr$ of $\cS^*_\infty[Q;Y]$ is given
by the expression
\be
\vr=\op\sum_{k>0} \frac1k\ol\vr\circ h_k\circ h^n, \qquad
\ol\vr(\f)= \op\sum_{0\leq|\La|} (-1)^{\nm\La}\th^A\w
[d_\La(\dr^\La_A\rfloor\f)], \qquad \f\in \cS^{>0,n}_\infty[Q;Y],
\ee
similar to (\ref{r12}). The graded variational operator
$\dl=\vr\circ d$ is introduced. Then the BGDA $\cS^*_\infty[Q;Y]$
is split into the $\Bbb Z_2$-graded variational bicomplex
analogous to the above mentioned variational bicomplex of
$\cO^*_\infty Y$.

We restrict our consideration to the short variational subcomplex
\mar{g111}\beq
0\ar \Bbb R\ar \cS^0_\infty[Q;Y]\ar^{d_H}\cS^{0,1}_\infty[Q;Y]
\cdots \ar^{d_H} \cS^{0,n}_\infty[Q;Y]\ar^\dl \bE_1 \label{g111}
\eeq
and the subcomplex of one-contact graded forms
\mar{g112}\beq
 0\to \cS^{1,0}_\infty[Q;Y]\ar^{d_H} \cS^{1,1}_\infty[Q;Y]\cdots
\ar^{d_H}\cS^{1,n}_\infty[Q;Y]\ar^\vr \bE_1\to 0, \label{g112}
\eeq
of the BGDA $\cS^*_\infty[Q;Y]$. They possess the following
cohomology.$^4$

\begin{theo} \label{g96} \mar{g96} The cohomology of
the complex (\ref{g111}) equals the de Rham cohomology $H^*(X)$ of
$X$.
\end{theo}

\begin{theo} \label{g96'} \mar{g96'}
The complex (\ref{g112}) is exact.
\end{theo}

One can think of the elements
\mar{0709}\beq
L=\cL\om\in \cS^{0,n}_\infty[Q;Y], \qquad \dl L= \th^A\w
\cE_A\om=\op\sum_{0\leq|\La|}
 (-1)^{|\La|}\th^A\w d_\La (\dr^\La_A L)\om\in \bE_1 \label{0709}
\eeq
of the complexes (\ref{g111}) -- (\ref{g112}) as being a {\it
graded Lagrangian} and its Euler--Lagrange operator, respectively.
The components $\cE_A$ of $\dl L$ are graded variational
derivatives.

By virtue of Theorem \ref{g96}, every $\dl$-closed graded
Lagrangian $L$ (\ref{0709}) is the sum
\mar{g215}\beq
\f=\psi + d_H\xi, \qquad \xi\in \cS^{0,n-1}_\infty[Q;Y],
\label{g215}
\eeq
where $\psi$ is a non-exact $n$-form on $X$.

The global exactness of the complex (\ref{g112}) at the term
$\cS^{1,n}_\infty[Q;Y]$ results in the following.$^4$

\begin{prop} \label{g103} \mar{g103}
Given a graded Lagrangian $L=\cL\om$, there is the decomposition
\mar{g99,'}\ben
&& dL=\dl L - d_H\Xi,
\qquad \Xi\in \cS^{1,n-1}_\infty[Q;Y], \label{g99}\\
&& \Xi=\op\sum_{s=0} \th^A_{\nu_s\ldots\nu_1}\w
F^{\la\nu_s\ldots\nu_1}_A\om_\la,\qquad F_A^{\nu_k\ldots\nu_1}=
\dr_A^{\nu_k\ldots\nu_1}\cL-d_\la F_A^{\la\nu_k\ldots\nu_1}
+h_A^{\nu_k\ldots\nu_1},  \label{g99'}
\een
where local graded functions $h$ obey the relations $h^\nu_a=0$,
$h_a^{(\nu_k\nu_{k-1})\ldots\nu_1}=0$.
\end{prop}

Proposition \ref{g103} states the existence of a global finite
order Lepagean equivalent $\Xi_L=\Xi+L$ of any graded Lagrangian
$L$. Locally, one can always choose $\Xi$ (\ref{g99'}) where all
functions $h$ vanish.

\bigskip
\bigskip

\noindent {\bf VI BRST SYMMETRIES}
\bigskip

A graded derivation  $\vt\in\gd \cS^0_\infty[Q;Y]$ of the $\Bbb
R$-ring $\cS^0_\infty[Q;Y]$ is said to be contact if the Lie
derivative $\bL_\vt$ preserves the ideal of contact graded forms
of the BGDA $\cS^*_\infty[Q;Y]$. With respect to the local basis
$\{s^A\}$ for the BGDA $\cS^*_\infty[Q;Y]$, any contact graded
derivation takes the form
\mar{g105}\beq
\vt=\vt_H+\vt_V=\vt^\la d_\la + (\vt^A\dr_A +\op\sum_{|\La|>0}
d_\La\vt^A\dr_A^\La), \label{g105}
\eeq
where tuple of graded derivations $\{\dr_\la,\dr^\La_A\}$ is
defined as the dual of the tuple $\{dx^\la, ds^A_\La\}$ of
generating elements of the $\cS^0_\infty[Q;Y]$-algebra
$\cS^*_\infty[Q;Y]$, and $\vt^\la$, $\vt^A$ are local graded
functions.$^4$ The interior product $\vt\rfloor\f$ and the Lie
derivative $\bL_\vt\f$, $\f\in\cS^*_\infty[Q;Y]$, are defined by
the same formulae
\be
&& \vt\rfloor \f=\vt^\la\f_\la + (-1)^{[\f_A]}\vt^A\f_A, \qquad
\f\in \cS^1_\infty[Q;Y],\\
&& \vt\rfloor(\f\w\si)=(\vt\rfloor \f)\w\si
+(-1)^{|\f|+[\f][\vt]}\f\w(\vt\rfloor\si), \qquad \f,\si\in
\cS^*_\infty[Q;Y], \\
&& \bL_\vt\f=\vt\rfloor d\f+ d(\vt\rfloor\f), \qquad
\bL_\vt(\f\w\si)=\bL_\vt(\f)\w\si
+(-1)^{[\vt][\f]}\f\w\bL_\vt(\si),
\ee
as those on a graded manifold. One can justify that any vertical
contact graded derivation $\vt$ (\ref{g105}) satisfies the
relations
\mar{g232}\beq
\vt\rfloor d_H\f=-d_H(\vt\rfloor\f), \qquad
\bL_\vt(d_H\f)=d_H(\bL_\vt\f), \qquad \f\in\cS^*_\infty[Q;Y].
\label{g232}
\eeq

\begin{prop}  \label{g106} \mar{g106}
It follows from the splitting (\ref{g103}) that the Lie derivative
$\bL_\vt L$ of a Lagrangian $L$ along a contact graded derivation
$\vt$ (\ref{g105}) fulfills the first variational formula
\mar{g107}\beq
\bL_\vt L= \vt_V\rfloor\dl L +d_H(h_0(\vt\rfloor \Xi_L)) + d_V
(\vt_H\rfloor\om)\cL, \label{g107}
\eeq
where $\Xi_L=\Xi+L$ is a Lepagean equivalent of $L$ given by the
coordinate expression (\ref{g99'}).$^4$
\end{prop}

A contact graded derivation $\vt$ is said to be variational if the
Lie derivative (\ref{g107}) is $d_H$-exact. A glance at the
expression (\ref{g107}) shows that: (i) a contact graded
derivation $\vt$ is variational only if it is projected onto $X$,
(ii) $\vt$ is variational iff its vertical part $\vt_V$ is
variational.

Therefore, we restrict our consideration to vertical contact
graded derivations
\mar{0672}\beq
\vt=\op\sum_{0\leq|\La|} d_\La\up^A\dr_A^\La, \label{0672}
\eeq
where the tuple of graded derivations $\{\dr^\La_A\}$ is defined
as the dual of the tuple $\{\th^A_\La\}$ of contact graded forms.
Such a derivation is completely determined by its first summand
\mar{0673}\beq
\up=\up^A(x^\la,s^A_\La)\dr_A, \qquad 0\leq|\La|\leq k,
\label{0673}
\eeq
which is also a graded derivation of $\cS^0_\infty[Q;Y]$. It is
called the {\it generalized graded vector field}. A glance at the
first variational formula (\ref{g107}) shows that $\vt$
(\ref{0672}) is variational iff $\up\rfloor \dl L$ is $d_H$-exact.

A vertical contact graded derivation $\vt$ (\ref{0672}) is said to
be nilpotent if
\mar{g133}\beq
\bL_\up(\bL_\up\f)= \op\sum_{|\Si|\geq 0,|\La|\geq 0 }
(\up^B_\Si\dr^\Si_B(\up^A_\La)\dr^\La_A +
(-1)^{[s^B][\up^A]}\up^B_\Si\up^A_\La\dr^\Si_B \dr^\La_A)\f=0
\label{g133}
\eeq
for any horizontal graded form $\f\in \cS^{0,*}_\infty[Q;Y]$. One
can show$^4$ that $\vt$ is nilpotent only if it is odd and iff the
equality
\mar{0688}\beq
\bL_\vt(\up^A)=\op\sum_{|\Si|\geq 0} \up^B_\Si\dr^\Si_B(\up^A)=0
\label{0688}
\eeq
holds for all $\up^A$.

\begin{ex} \label{0675} \mar{0675} Let $Y\to X$ be an affine
bundle, $L$ (\ref{0512}) a Lagrangian of a gauge theory on $Y$ and
$\up$ (\ref{0509}) its gauge symmetry where $E=Y\op\times_X V$ is
the pull-back onto $Y$ of a vector bundle $V\to X$ coordinated by
$(x^\la,\xi^r)$. Let us consider the BGDA
$\cS^*_\infty[V;Y]=\cC^*_\infty V\w\cP^*_\infty Y$ possessing a
local basis $\{c^r, y^i\}$. Let $L\in \cO^{0,n}_\infty Y$ be a
polynomial in $y^i_\La$, $0\leq |L|$. Then it is a graded
Lagrangian $L\in \cP^{0,n}_\infty Y\subset \cS^{0,n}_\infty[V;Y]$
in $\cS^*_\infty[V;Y]$.  Since $E\to Y$ is the pull-back bundle, a
gauge symmetry $\up$ (\ref{0509}) gives rise to the generalized
vector field $\up_E=\up$ on $E$, and the latter defines the
generalized graded vector field $\up$ (\ref{0673}) by the formula
\mar{0680}\beq
\up= \op\sum_{0\leq|\Xi|\leq m}\up^{i,\Xi}_r(x^\la,y^i_\Si)c^r_\Xi
\dr_i. \label{0680}
\eeq
It is easily justified that the vertical contact graded derivation
$\vt$ (\ref{0672}) generated by $\up$ (\ref{0680}) is variational
for the graded Lagrangian $L$. It is odd, but need not be
nilpotent. However, one can try to find a nilpotent contact graded
derivation generated by some generalized graded vector field
\mar{0684}\beq
\up= \op\sum_{0\leq|\La|\leq m}\up^{i,\La}_r c^r_\La \dr_i +
u^r\dr_r \label{0684}
\eeq
which coincides with $\vt$ on $\cP^*_\infty Y$. In this case, the
nilpotency conditions (\ref{0688}) read
\mar{0690,1}\ben
&& \op\sum_\Si d_\Si(\op\sum_\Xi\up^{i,\Xi}_rc^r_\Xi)
\op\sum_\La\dr^\Si_i (\up^{j,\La}_s)c^s_\La +\op\sum_\La d_\La
(u^r)\up^{j,\La}_r
=0, \label{0690}\\
&& \op\sum_\La(\op\sum_\Xi d_\La(\up^{i,\Xi}_r c^r_\Xi)\dr^\La_i
+d_\La (u^r)\dr_r^\La)u^q=0\label{0691}
\een
for all indices $j$ and $q$. They are equations for graded
functions $u^r\in\cS^0_\infty[V;Y]$. Since these functions are
polynomials
\mar{0693}\beq
u^r=u_{(0)}^r + \op\sum_\G u_{(1)p}^{r,\G} c^p_\G +
\op\sum_{\G_1,\G_2} u_{(2)p_1p_2}^{r,\G_1\G_2}
c^{p_1}_{\G_1}c^{p_2}_{\G_2} +\cdots \label{0693}
\eeq
in $c^s_\La$, the equations (\ref{0690}) -- (\ref{0691}) take the
form
\mar{0694}\ben
&& \op\sum_\Si d_\Si(\op\sum_\Xi\up^{i,\Xi}_rc^r_\Xi)
\op\sum_\La\dr^\Si_i (\up^{j,\La}_s)c^s_\La +\op\sum_\La d_\La
(u_{(2)}^r)\up^{j,\La}_r
=0, \label{0694a}\\
&& \op\sum_\La d_\La (u_{(k\neq 2)}^r)\up^{j,\La}_r =0,
\label{0694b}\\
&& \op\sum_\La\op\sum_\Xi d_\La(\up^{i,\Xi}_r c^r_\Xi)\dr^\La_i
u_{(k-1)}^q +\op\sum_{m+n-1=k}d_\La (u_{(m)}^r)\dr_r^\La u_{(n)}^q
=0. \label{0694c}
\een
One can think of the equalities (\ref{0694a}) and (\ref{0694c})
(and, consequently, the nilpotency conditions (\ref{0690}) --
(\ref{0691})) as being the generalized commutation relations and
generalized Jacobi identities of gauge transformations,
respectively.$^{17}$ For instance, let us consider a gauge system
on a principal bundle and the generalized vector field $\up$
(\ref{0660}) in Example \ref{0650}. Following the procedure above,
we replace parameters $\xi^r$ and $\tau^\la$ with the odd ghosts
$C^r$ and $C^\la$, respectively, and obtain the generalized graded
vector field
\mar{0695}\beq
\up= (c^r_{pq}a^p_\la C^q + C^r_\la -a^r_\m C^\m_\la-C^\m
a_{\m\la}^r)\dr^\la_r +(-\frac12c^r_{pq}C^pC^q -C^\m C^r_\m)\dr_r
+ C^\la_\m C^\m\dr_\la \label{0695}
\eeq
such that the vertical contact graded derivation (\ref{0672})
generated by $\up$ (\ref{0695}) is nilpotent. In the case of the
vertical gauge symmetry (\ref{0662}), we obtain the familiar BRST
transformation
\mar{0701}\beq
\up= (c^r_{pq}a^p_\la C^q + C^r_\la)\dr^\la_r
-\frac12c^r_{pq}C^pC^q\dr_r  \label{0701}
\eeq
of Yang--Mills theory.
\end{ex}

Generalizing Example \ref{0675}, we describe BRST symmetries in a
general setting as follows.

Let $\cS^*_\infty[Q;Y]$ be the BGDA (\ref{0670}) and $L\in
\cS^{0,n}_\infty[Q;Y]$ a graded Lagrangian. We agree to call
generating elements $s^A$ of $\cS^*_\infty[Q;Y]$ the fields. Let
$V\to X$ be a vector bundle coordinated by $(x^\la, \xi^r)$. By
analogy with $\cS^*_\infty[Q;Y]$, we consider the BGDA
\mar{0702}\beq
\cS^*_\infty[QV;Y]=\cC^*_\infty[Q\op\times_X V]\w \cP^*_\infty Y,
\label{0702}
\eeq
whose local basis is $\{s^A, c^r\}$. Obviously, $L$ is also a
graded Lagrangian in $\cS^*_\infty[QV;Y]$. Let
\mar{0703}\beq
\vt=\op\sum_{0\leq|\La|} (d_\La\up^A\dr_A^\La
+d_\La\up^r\dr_r^\La) \label{0703}
\eeq
be a graded contact derivation of the $\Bbb R$-ring
$\cS^0_\infty[QV;Y]$ generated by an odd generalized graded vector
field
\mar{0704}\beq
\up=\up^A\dr_A +\up^r\dr_r \label{0704}
\eeq
whose restriction to $\cS^*_\infty[Q;Y]$ is linear in $c^r_\La$,
i.e.,
\mar{0705}\beq
\up= \op\sum_{0\leq|\Xi|\leq m}c^r_\Xi\up^{A,\Xi}_r(x^\la,s^B_\Si)
\dr_A + \up^r\dr_r. \label{0705}
\eeq
If $\vt$ (\ref{0703}) is variational for $L$ and nilpotent, we say
that $\up$ (\ref{0705}) is a {\it BRST symmetry} of $L$. Following
the terminology of BRST theory, we agree to call generating
elements $c^r$ of $\cS^*[QV;Y]$ the {\it ghosts}.

\bigskip
\bigskip

\noindent {\bf VII. NOETHER'S SECOND THEOREM II}
\bigskip

In order to introduce Noether identities in the case of BRST
symmetries, let us extend the BGDA $\cS^*[QV;Y]$ (\ref{0702}) to
the BGDA
\mar{0706}\beq
\cS^*_\infty[Q\ol Y^*V;Y\ol Q^*]=\cC^*_\infty[Q\op\times_X \ol
Y^*\op\times_X V]\w \cP^*_\infty[Y\op\times_X \ol Q^*],
\label{0706}
\eeq
where $\ol Q^*$ is the density-dual of $Q$ and $\ol Y^*$ is the
density dual of the vector bundle $\wt Y\to X$ which the affine
bundle $Y$ is modelled on (e.g., $\wt Y=Y$ if $Y$ is a vector
bundle). The local basis for the BGDA $\cS^*_\infty[Q\ol Y^*V;Y\ol
Q^*]$ is $\{s^A, c^r, \ol s_A\}$. Following the terminology of the
field-antified BRST theory, we call generating elements $\ol s_A$
of $\cS^*_\infty[Q\ol Y^*V;Y\ol Q^*]$ the {\it antifields}. Their
Grassmann parity is $[\ol s_A]=([s^A]+1){\rm mod}\,2$.

The  BGDA $\cS^*_\infty[Q\ol Y^*V;Y\ol Q^*]$ (\ref{0706}) is
provided with the {\it Koszul--Tate differential} defined as the
nilpotent contact graded derivation
\mar{0707}\beq
\ol\dl=\op\sum_{0\leq|\La|}\rdr^{\La A}d_\La\cE_A, \label{0707}
\eeq
where $\cE_A$ are the graded variational derivatives (\ref{0709})
and the tuple of graded right derivations $\rdr^{\La A}$ is the
dual of the tuple of contact graded forms $\{\th_{\La A}\}$, i.e.,
\be
\th_{\La A}\lfloor \rdr^{\Si B}=\dl^\Si_\La \dl^A_B,
\ee
where multi-indices $\Si$ and $\La$ are regard modulo
permutations. Because of the expression (\ref{0709}) for $\dl L$,
it is convenient to describe the Koszul--Tate differential as a
graded derivation acting on graded functions and forms $\f$ on the
right by the rule
\be
\ol\dl(\f)=d\f\lfloor\ol\dl +d(\f\lfloor\ol\dl), \qquad
\ol\dl(\f\w\f')=(-1)^{[\f']}\ol\dl(\f)\w\f'+ \f\w\ol\dl(f').
\ee

\begin{defi} \label{0710} \mar{0710}
Given a graded Lagrangian $L\in \cS^*_\infty[Q;Y]\subset
\cS^*_\infty[Q\ol Y^*V;Y\ol Q^*]$ (\ref{0709}), we say that its
Euler--Lagrange operator $\dl L$ (\ref{0709}) obeys a Noether
identity if there exists a $\ol\dl$-closed even graded density
\mar{0712}\beq
\Delta= c^r\Delta_r\om=\op\sum_{0\leq|\La|\leq m}
c^r\Delta^{A,\La}_r(x^\la,s^B_\Si)\ol s_{\La A} \om \in
\cS^{0,n}_\infty[Q\ol Y^*V;Y\ol Q^*], \label{0712}
\eeq
which is linear both in ghosts $c^r$ and antifields $\ol s_A$ and
their jets $\ol s_{\La A}$. The above mentioned {\it Noether
identity} reads
\mar{0713}\beq
\ol\dl(\Delta)=c^r[\op\sum_{0\leq|\La|\leq m} \Delta^{A,\La}_r
d_\La \cE_A] \om=0. \label{0713}
\eeq
\end{defi}

Then Noether's second theorem for BRST symmetries is formulated as
follows.

\begin{theo} \label{0716} \mar{0716}
If $\up$ (\ref{0705}) is a BRST symmetry of a graded Lagrangian
$L$, then
\mar{0714}\ben
&& \Delta=\eta(\up)=\op\sum_{0\leq|\La|\leq m}
c^r\eta(\up)^{A,\La}_r\ol s_{\La A} \om, \label{0714}\\
&& \eta(\up)^{A,\La}_r=\op\sum_{0\leq|\Si|\leq
m-|\La|}(-1)^{|\Si+\La|}C^{|\Si|}_{|\Si+\La|} d_\Si
\up^{A,\Si+\La}_r, \nonumber
\een
is a $\ol\dl$-closed graded density (\ref{0712}). Conversely, if a
$\ol\dl$-closed graded density $\Delta$ (\ref{0712}) exists, the
generalized graded vector field
\mar{0715}\ben
&& \up= \eta(\Delta)=\op\sum_{0\leq|\Xi|\leq m}c^r_\Xi
\eta(\Delta)^{A,\Xi}_r \dr_A, \label{0715}\\
&& \eta(\Delta)^{A,\La}_r=\op\sum_{0\leq|\Si|\leq
m-|\La|}(-1)^{|\Si+\La|} C^{|\Si|}_{|\Si+\La|}d_\Si
\Delta^{A,\Si+\La}_r, \nonumber
\een
generates a contact graded derivation (\ref{0703}) which is
variational for the graded Lagrangian $L$, but it need not be
nilpotent. The relations (\ref{0634}) hold.
\end{theo}

\begin{proof}
 The first
summand of the generalized vector field (\ref{0705}) defines the
graded function
\mar{0810}\beq
 \up= \op\sum_{0\leq|\Xi|\leq
m}c^r_\Xi\up^{A,\Xi}_r(x^\la,s^B_\Si)s_A\in
\cS^*_\infty[QQ^*V;Y\wt Y^*], \label{0810}
\eeq
and {\it vice versa}. Then the proof follows from Theorem
\ref{0717} in Appendix B. By virtue of this theorem, the graded
function (\ref{0810}) yields the graded density (\ref{0714}).
Since the graded density $\up\rfloor\dl L$ is $d_H$-exact, we
obtain the equality
\be
\dl(\up\rfloor\dl L)=\ol\dl(\eta(\up))=0.
\ee
Conversely, the graded density (\ref{0712}) yields the graded
function (\ref{0810}) where
$\up^{A,\Xi}_r=\eta(\Delta)^{A,\Xi}_r$. Since $\Delta$
(\ref{0712}) is $\ol\dl$-closed, we have
\be
&&0=\op\sum_{0\leq|\La|\leq m}
c^r\Delta^{A,\La}_rd_\La\cE_A\om=\op\sum_{0\leq|\La|\leq m}
(-1)^{|\La|}d_\La(c^r \Delta^{A,\La}_r)\cE_A\om +
d_H\si=\\
&& \qquad \op\sum_{0\leq|\Xi|\leq m}c^r_\Xi\up^{A,\Xi}_r\cE_A\om
+d_H\si=\up\rfloor \dl L+d_H\si,
\ee
i.e., the graded contact derivation generated by $\up$
(\ref{0715}) is variational for $L$. Due to the equality
(\ref{0606d}), the relations (\ref{0634}) hold.
\end{proof}

Bearing in mind the field-antifield BRST theory and BV
quantization, the Noether identity (\ref{0713}) can be rewritten
as follows. Let us consider the BGDA $\cS^*_\infty[Q\ol Y^*V;Y\ol
Q^*\ol V^*]$ possessing the local basis $\{s^A, c^r, \ol s_A, \ol
c_r\}$, where even elements $\ol c_r$ are called {\it antighosts}
of the ghosts $c^r$. Clearly, the graded density $\Delta$
(\ref{0712}) is an element of
 $\cS^*_\infty[Q\ol Y^*V;Y\ol Q^*\ol V^*]$. Then this BGDA
is provided with the contact graded right derivation
\mar{0812}\beq
\ol\dl_c=\op\sum_{0\leq|\La|}(\rdr^{\La A} d_\La\cE_A+ \rdr^{\La
r} d_\La\Delta_r), \label{0812}
\eeq
where the tuple of graded right derivations $\rdr^{\La r}$ is the
dual of $\th_{\La r}$. It is easily justified that the graded
density $\Delta$ (\ref{0712}) obeys the Noether identity
(\ref{0713}) iff the graded right derivation $\ol\dl_c$
(\ref{0812}) is nilpotent. It is the extension of the Koszul--Tate
differential (\ref{0707}) to antighosts. For instance, the graded
density $\Delta$ is always $\ol\dl_c$-exact.

\bigskip
\bigskip

\noindent {\bf VIII. REDUCIBLE BRST SYMMETRIES}
\bigskip

The notion of a reducible Noether identity in Section IV is
straightforwardly generalized to BRST symmetries, but we formulate
it in terms of the Koszul--Tate differential. We say that the
Noether identity (\ref{0713}) is $N$-stage reducible if the
following conditions hold.

(a) There exists a set of vector bundles $V_{-1}=V,V_0, \ldots,
V_N$ over $X$, and we consider the BGDA
\mar{0820}\ben
&& \ol\cS^*_\infty\{N\}=\label{0820}\\
&& \qquad\cS^*_\infty[Q\ol Y^*VV_1\ldots V_{2k-1}\ldots \ol
V_0^*\ldots\ol V_{2k}^*\ldots;Y\ol Q^*V_0\ldots V_{2k}\ldots \ol
V^*\ol V^*_1\ldots \ol V_{2k-1}^*\ldots]. \nonumber
\een
It possesses a local basis
\mar{0821}\beq
\{s^A,\ol s_A, c^r, c^{r_0}, \ldots, c^{r_N}, \ol c_r, \ol
c_{r_0}, \ldots, \ol c_{r_N}\}, \qquad [c^{r_k}]=k\,{\rm
mod}\,2,\label{0821}
\eeq
where $c^{r_k}$ and $\ol c_{r_k}$ are called the $k$-stage ghosts
and antighosts, respectively..

 (b) The BGDA (\ref{0820}) contains the graded density $\Delta$ (\ref{0712}) and
 a set of even graded densities
 \mar{0822}\beq
\Delta_{(k)}=c^{r_k}\Delta_{r_k}\om=\op\sum_\La
c^{r_k}\Delta_{r_k}^{r_{k-1},\La}(x^\la, s^B_\Si)\ol c_{\La
r_{k-1}}\om, \qquad k=0,\ldots, N, \label{0822}
 \eeq
such that the contact graded derivation
\mar{0823}\beq
\ol\dl_N=\op\sum_{0\leq|\La|} (\rdr^{\La A} d_\La\cE_A+ \rdr^{\La
r} d_\La\Delta_r + \rdr^{\La r_0} d_\La\Delta_{r_0}+ \cdots +
\rdr^{\La r_N} d_\La\Delta_{r_N}) \label{0823}
\eeq
is weakly nilpotent, i.e., $\ol\dl_N(\ol\dl_N(f))$ is
$\ol\dl$-exact for any graded function $f\in
\ol\cS^0_\infty\{N\}$. This nilpotency condition is equivalent to
the requirement that all the compositions
\mar{0826}\beq
\op\sum_\Xi c^{r_k}\Delta^{r_{k-1},\Xi}_{r_k}d_\Xi
(\op\sum_\La\Delta^{r_{k-2},\La}_{r_{k-1}}\ol c_{r_{k-2}}), \qquad
k=1, \ldots, N, \label{0826}
\eeq
are $\ol\dl$-exact. The graded derivation $\ol\dl_N$ (\ref{0823})
is called the {\it $N$-stage Koszul--Tate differential}.

(c) No graded density $\Delta$, $\Delta_{(k)}$, $k=0,\ldots,N$, is
$\ol\dl$-exact. Let $V'_0, \ldots, V'_{N'}$ be another set of
vector bundles  such that $V'_{k\leq N}$ contains $V_k$ as a
direct summand. Then the corresponding BGDA
$\ol\cS^*_\infty\{N'\}$ (\ref{0820}) contains the graded densities
$\Delta$ (\ref{0712}), $\Delta_{(k)}$ (\ref{0822}), $k=0,\ldots,
N$, and it is provided with the contact graded derivation
$\ol\dl_N$ (\ref{0823}). If there exists another set
$\Delta'_{(k)}$, $k=0,\ldots, N'$, of graded densities obeying the
conditions in item (b), then any graded density $\Delta'_{(k)}$ of
this set is $\ol\dl_N$-exact.

Note, that following the arguments in Section III, one can say
that the Noether identity (\ref{0713}) is complete if it obeys the
condition (c).

If the Noether identity (\ref{0713}) is reducible, the associated
BRST symmetry (\ref{0715}) is reducible as follows.

Let us consider the BGDA
\mar{0828}\beq
\cS^*_\infty\{N\}=\cS^*_\infty[QQ^*VV^*V_1V_1^*\ldots
V_{2k-1}V^*_{2k-1}\ldots ;Y\wt Y^*V_0V_0^*\ldots
V_{2k}V^*_{2k}\ldots], \label{0828}
\eeq
possessing the local basis $\{s^A,s_A, c^r, c^{r_0}, \ldots,
c^{r_N},c_r, c_{r_0}, \ldots, c_{r_N}\}$. By virtue of Theorem
\ref{0717}, each graded density $\Delta_{(k)}$ (\ref{0822}),
$k=0,\ldots, N$, defines the graded function
\be
\up_{(k)} =\op\sum_\Xi
c^{r_k}_\Xi\eta(\Delta_{(k)})^{r_{k-1},\Xi}_{r_{k}}c_{r_{k-1}}\in
\cS^0_\infty(\{N\}
\ee
and, consequently, the generalized graded vector field
\mar{0829}\beq
\up_{(k)} =\op\sum_\Xi
c^{r_k}_\Xi\eta(\Delta_{(k)})^{r_{k-1},\Xi}_{r_{k}}\dr_{r_{k-1}}\in
\cS^0_\infty(\{N\}, \label{0829}
\eeq
which yields a contact graded derivation of the BGDA
$\ol\cS^*_\infty\{N\}$. Similarly to the proof of Theorem
\ref{0580}, one can show that they possess the following
properties.

(a') Contact graded derivations $\vt$ and $\vt_{(k)}$ generated by
the generalized graded vector fields $\up$ (\ref{0715}) and
$\up_{(k)}$ (\ref{0829}) are not weakly $\ol\dl$-exact, i.e.,
$\vt(f)$ and $\vt_{(k)}(f)$ are not $\ol\dl$-exact for some graded
function $f\in \ol\cS^*_\infty\{N\}$.

(b') Contact graded derivations generated by the generalized
graded vector fields
\mar{0830,1}\ben
&& \op\sum_\La d_\La(\op\sum_\Xi
c^{r_0}_\Xi\eta(\Delta_{(0)})^{r,\Xi}_{r_{0}})
\eta(\Delta)^{A,\La}_r\dr_A, \label{0830}\\
&& \op\sum_\La d_\La(\op\sum_\Xi
c^{r_k}_\Xi\eta(\Delta_{(k)})^{r_{k-1},\Xi}_{r_{k}})
\eta(\Delta_{(k-1)})^{r_{k-2},\La}_{r_{k-1}}\dr_{r_{k-2}}, \qquad
k=1,\ldots, N, \label{0831}
\een
are weakly $\ol\dl$-exact. This condition can be reformulated as
follows. Let us consider the BGDA
\be
\cS^*_\infty[QQ\ol Y^*VV_1\ldots V_{2k-1}\ldots \ol V_0^*\ldots\ol
V_{2k}^*\ldots;Y\wt Y\ol Q^*V_0\ldots V_{2k}\ldots \ol V^*\ol
V^*_1\ldots \ol V_{2k-1}^*\ldots]
\ee
whose basis consists of elements (\ref{0821}) and the elements
$s'^A$ associated to the additional bundles $Q$ and $\wt Y$. Let
us replace the generalized graded vector field $\up$ (\ref{0715})
with
\mar{0833}\beq
\up'=\op\sum_{0\leq|\Xi|\leq m}c^r_\Xi \eta(\Delta)^{A,\Xi}_r
(x^\la, s^B_\Si) \dr'_A, \label{0833}
\eeq
where the graded derivations $\dr'_A$ are dual of the basis
elements $s'^A$. Then the contact graded derivation generated by
the generalized graded vector field
\mar{0832}\beq
\up_N=\up' +\up_{(0)}+\cdots + \up_{(N)} \label{0832}
\eeq
is weakly nilpotent.

(c') Let $\up_{N'}$ be the generalized graded vector field
(\ref{0832}) defined by the graded densities $\Delta'_{(k)}$ in
item (c) above. Then there exists some generalized graded vector
field $u$ such that $\up_{N'}-[u,\up_N]$ is weakly $\ol\dl$-exact.

Conversely, one can show the following. Let $\up$ (\ref{0715}) be
a BRST symmetry of a graded Lagrangian $L$. Let us assume that
there exists a set of generalized graded vector fields
\mar{0834}\beq
\up_{(k)} =\op\sum_\Xi
c^{r_k}_\Xi\up^{r_{k-1},\Xi}_{r_k}\dr_{r_{k-1}}\in \cS^0_\infty
\{N\}, \label{0834}
\eeq
which obey the conditions in items (a') -- (c'). Then the graded
density (\ref{0716}) defines a complete reducible Noether identity
where the graded densities
\be
\Delta_{(k)}=\op\sum_\La
c^{r_k}\eta(\up_{(k)})_{r_k}^{r_{k-1},\La}(x^\la, s^B_\Si)\ol
c_{\La r_{k-1}}
\ee
obey the conditions in items (a) -- (c).

In contrast with the generalized graded vector field $\up_N$
(\ref{0832}), the contact graded derivation generated by the
generalized graded vector field $\up +\up_{(0)}+\cdots +
\up_{(N)}$ need not be weakly nilpotent. Its extension to the
nilpotent one provides a BRST symmetry of the field-antifield BRST
theory whose Lagrangian depends on ghosts and antifields.

\bigskip
\bigskip

\noindent {\bf APPENDIX A}
\bigskip

A $k$-order differential operator on a fiber bundle $Y\to X$ with
values in a fiber bundle $Z\to X$ is defined as a section $\Delta$
of the fiber bundle $J^kY\op\times_X Z\to J^kY$. It admits an
$m$-order jet prolongation $\Delta^{(m)}$ as a section of the
fiber bundle
\be
J^{m+k}Y\op\times_X J^mZ\to J^{m+k}Y.
\ee
By a differential operator throughout is meant its appropriate
finite order jet prolongation. Given bundle coordinates
$(x^\la,y^i)$ on $Y$ and $(x^\la, z^A)$ on $Z$, a differential
operator $\Delta$  reads
\be
z^A\circ \Delta=\Delta^A(x^\la,y^i_\La), \qquad z^A_\Si\circ
\Delta^{(m)}=d_\Si\Delta^A,\qquad 0\leq|\La|\leq k, \qquad
0\leq|\Si|\leq m.
\ee

If $Z$ is a composite fiber bundle $\pi\circ\pi_{ZY}:Z\to Y\to X$
and the relation $\pi_{ZY}\circ\Delta=\pi^k_0$ holds, a
differential operator $\Delta$ is identified to a section of the
fiber bundle $J^kY\op\times_Y Z\to J^kY$ or, equivalently, a
bundle morphism $J^kY\ar_Y Z$.

Let $E\to Y$ and $Q\to Y$ be vector bundles. A $k$-order
$Q$-valued differential operator $\up$ on $E\to X$ is called
linear on $E\to Y$ (or, simply, linear) if $\up: J^kE\to Q$ is a
morphism of the vector bundle $J^kE\to J^kY$ to the vector bundle
$Q\to Y$ over $\pi^k_0:J^kY\to Y$. Given bundle coordinates
 $(x^\la,y^i,\xi^r)$ on $E$ and $(x^\la,y^i,q^a)$ on $Q$, such an
operator is represented by the function
\mar{0614}\beq
\up=\up^aq_a=\op\sum_{0\leq|\La|\leq m}
\up^{a,\La}_r(x^\la,y^i_\Si) \xi^r_\La q_a\in
\cO^0_\infty[E\op\times_Y Q^*], \qquad 0\leq|\Si|\leq m.
\label{0614}
\eeq

Let us consider the density-dual $\ol E^*$ (\ref{0630}) of a
vector bundle $E\to Y$ and that $\ol Q^*$ of $Q\to Y$ coordinated
by $(x^\la, y^i, \ol q_a)$. Let
 $\Delta$ be a linear $\ol E^*$-valued differential
operator on $\ol Q^*$. It is represented by the density
\mar{0615}\beq
\Delta=\Delta_r\xi^r\om=\op\sum_{0\leq|\La|\leq m}
\Delta^{a,\La}_r(x^\la,y^i_\Si)\ol q_{\La a}\xi^r\om\in
\cO^{0,n}_\infty[E\op\times_Y Q^*], \qquad 0\leq|\Si|\leq m.
\label{0615}
\eeq

\begin{theo} \label{0610} \mar{0610}
Any linear $Q$-valued differential operator $\up$ (\ref{0614}) on
$E$ yields the linear $\ol E^*$-valued differential operator
\mar{0616}\ben
&& \eta(\up)=\op\sum_{0\leq|\La|\leq m}
\eta(\up)^{a,\La}_r \ol q_{\La a}\xi^r\om, \label{0616}\\
&&\eta(\up)^{a,\La}_r=\op\sum_{0\leq|\Si|\leq
m-|\La|}(-1)^{|\Si+\La|}C^{|\Si|}_{|\Si+\La|} d_\Si
(\up^{a,\Si+\La}_r), \nonumber
\een
on $\ol Q^*$.  Conversely, any linear $\ol E^*$-valued
differential operator $\Delta$ (\ref{0615}) on $\ol Q^*$ defines
the linear $Q$-valued  differential operator
\mar{0617}\ben
&&\eta(\Delta)=\op\sum_{0\leq|\La|\leq m} \eta(\Delta)^{a,\La}_r
\xi^r_\La q_a, \label{0617}\\
&&\eta(\Delta)^{a,\La}_r=\op\sum_{0\leq|\Si|\leq
m-|\La|}(-1)^{|\Si+\La|}C^{|\Si|}_{|\Si+\La|} d_\Si
(\Delta^{a,\Si+\La}_r), \nonumber
\een
on $E$. The relations (\ref{0634}) hold.
\end{theo}

\begin{proof}  The function $\up$ (\ref{0614})
defines the  density
\mar{0637}\beq
\ol\up=\op\sum_{0\leq|\La|\leq m} \up^{a,\La}_r \xi^r_\La \ol
q_a\om\in \cO^{0,n}_\infty[E\op\times_Y Q^*].\label{0637}
\eeq
Its Euler--Lagrange operator
\be
\dl(\ol\up)=\cE_i dy^i\w\om +\cE_r d\xi^r\w\om + \cE^ad\ol
q_a\w\om
\ee
takes its values in the fiber bundle
\mar{0631}\beq
V^*(E\op\times_Y Q^*)\op\ot_{E\op\times_Y Q^*}\op\w^n T^*X,
\label{0631}
\eeq
where $V^*(E\op\times_Y Q^*)$ is the vertical cotangent bundle of
the fiber bundle $E\op\times_Y Q^*\to X$. Let
\mar{0633}\beq
\al_E: V^*(E\op\times_Y Q^*)\to V^*_Y(E\op\times_Y Q^*)\to V^*_YE
\label{0633}
\eeq
be the canonical projection of $V^*(E\op\times_Y Q^*)$ onto the
vertical cotangent bundle $V^*_Y(E\op\times_Y Q^*)$ of the fiber
bundle $E\op\times_Y Q^*\to Y$ and, afterwards, onto the vertical
cotangent bundle $V^*_YE$ of $E\to Y$. Then we obtain a
differential operator $(\al_E\circ\dl)(\ol\up)$ on $E\op\times_Y
Q^*$ with values in the fiber bundle $V^*_YE\op\ot_E\op\w^n T^*X$.
It reads
\be
(\al_E\circ\dl)(\ol\up)=\cE_r\ol
d\xi^r\ot\om=\op\sum_{0\leq|\La|\leq m}
(-1)^{|\La|}d_\La(\up^{a,\La}_r \ol q_a)\ol d\xi^r\ot\om,
\ee
where $\{\ol d\xi^r\}$ is the fiber basis for $V^*_YE\to E$. Due
to the canonical isomorphism $V^*_YE=E^*\op\times_Y E$, this
operator defines the  density
\be
\op\sum_{0\leq|\La|\leq m} (-1)^{|\La|} d_\La(\up^{a,\La}_r \ol
q_a) \xi^r\om \in\cO^{0,n}_\infty[E\op\times_Y Q^*]
\ee
and, by virtue of the formula (\ref{0606b}), the desired
differential operator (\ref{0616}). Conversely,  the
Euler--Lagrange operator of the density (\ref{0615}) takes its
values in the fiber bundle (\ref{0631}) and reads
\mar{0632}\beq
\dl(\ol\up)=\cE_i dy^i\w\om +\cE_r d\xi^r\w\om + \cE^a d\ol
q_a\w\om. \label{0632}
\eeq
In  order to repeat the above mentioned procedure, let us consider
a volume form $J\om$ on $X$ and substitute $d\ol
q_a\w\om=Jdq_a\w\om$ into the expression (\ref{0632}). Using the
projection
\be
\al_Q: V^*(E\op\times_Y Q^*)\to V^*_YQ^*
\ee
similar to $\al_E$ (\ref{0633}) and the canonical isomorphism
$V^*_YQ^*=Q\op\times_Y Q^*$, we come to the density
\be
\op\sum_{0\leq|\La|\leq m} (-1)^{|\La|} d_\La(\Delta^{a,\La}_r
\xi^r) q_a J\om \in\cO^{0,n}_\infty[E\op\times_Y Q^*]
\ee
and, hence, the function
\be
\op\sum_{0\leq|\La|\leq m} (-1)^{|\La|} d_\La(\Delta^{a,\La}_r
\xi^r) q_a  \in\cO^0_\infty[E\op\times_Y Q^*],
\ee
defining the desired operator (\ref{0617}). The relations
(\ref{0634}) result from the relation (\ref{0606d}).
\end{proof}

Relations (\ref{0634}) show that the intertwining operator $\eta$
(\ref{0616}) -- (\ref{0617}) provides a  bijection between the
sets Diff$(E,Q)$ and Diff$(\ol Q^*,\ol E^*)$ of differential
operators (\ref{0614}) and (\ref{0615}).

\begin{prop} \label{0621} \mar{0621}
Compositions of operators $\up\circ\up'$ and $\Delta'\circ \Delta$
obey the relations (\ref{0622}).
\end{prop}

\begin{proof}
It suffices to prove the first relation. Let $\up\circ\up'\in{\rm
Diff}(E',Q)$ be a composition of differential operators
$\up\in{\rm Diff}(E,Q)$ and $\up'\in{\rm Diff}(E',E)$. Given fiber
coordinates $(\xi^r)$ on $E\to Y$, $(\e^p)$ on $E'\to Y$ and $(\ol
q_a)$ on $\ol Q^*\to Y$, this composition defines the density
(\ref{0637})
\be
\ol{\up\circ\up'}=\op\sum_\La \up^{a,\La}_r
d_\La(\op\sum_\Si\up'^{r,\Si}_p\e^p_\Si)\ol q_a\om.
\ee
Following the relation (\ref{0606a}), one can bring this density
into the form
\be
\op\sum_\Si\up'^{r,\Si}_p\e^p_\Si\op\sum_\La (-1)^{|\La|} d_\La
(\up^{a,\La}_r \ol q_a)\om + d_H\si=
\op\sum_\Si\up'^{r,\Si}_p\e^p_\Si\op\sum_\La\eta(\up)^{a,\La}_r
\ol q_{\La a}\om +d_H\si.
\ee
Its Euler--Lagrange operator projected to $V^*_YE'\op\ot_{E'}
T^*X$ is
\be
\op\sum_\Si (-1)^{|\Si|} d_\Si(\up'^{r,\Si}_p
\op\sum_\La\eta(\up)^{a,\La}_r \ol q_{\La a})\ol
d\e^p\ot\om=\op\sum_\Si\eta(\up')^{r,\Si}_p
d_\Si(\op\sum_\La\eta(\up)^{a,\La}_r \ol q_{\La a})\ol
d\e^p\ot\om,
\ee
that leads to the desired composition $\eta(\up')\circ\eta(\up)$.
\end{proof}

\bigskip
\bigskip

\noindent {\bf APPENDIX B}
\bigskip

The following is a graded counterpart of Theorem \ref{0610}.

Let $T\to X$ and $W\to X$ be vector bundles, $W^*$ the dual of
$W$, and $\ol W^*$ the density-dual (\ref{0630}) of $W$. Given the
BGDA $\cS^*_\infty[Q;Y]$ (\ref{0670}), let us consider its
extensions to a BGDA $\cS^*_\infty[T,W^*]$ with the local basis
$\{s^A,t^r, w_a\}$, where elements $t^r$ and $w_a$ are either even
or odd, and to a BGDA $\cS^*_\infty[T,\ol W^*]$ possessing the
local basis $\{s^A,t^r,\ol w_a\}$, where $[\ol w_a]=([w_a]+1){\rm
mod}\,2$.

\begin{theo} \label{0717} \mar{0717}
Given a graded function
\mar{0801}\beq
\up=\op\sum_{0\leq|\La|\leq m} t^r_\La\up^{a,\La}_r(x^\la,s^A_\Si)
 w_a\in \cS^0_\infty[T,W^*], \label{0801}
\eeq
linear in $t^r_\La$ and $w_a$, there exists a graded density
\mar{0802}\ben
&&\eta(\up)=\op\sum_{0\leq|\La|\leq m}
t^r\eta(\up)^{a,\La}_r \ol w_{\La a}\om\in \cS^{0,n}_\infty[T,\ol W^*], \label{0802}\\
&&\eta(\up)^{a,\La}_r=\op\sum_{0\leq|\Si|\leq
m-|\La|}(-1)^{|\Si+\La|}C^{|\Si|}_{|\Si+\La|} d_\Si
(\up^{a,\Si+\La}_r), \nonumber
\een
linear in $t^r$ and $\ol w_{\La a}$. Conversely, such a density
\mar{0803}\beq
\Delta=\op\sum_{0\leq|\La|\leq m} t^r\Delta^{a,\La}_r\ol w_{\La
a}\om \label{0803}
\eeq
defines the graded function
\mar{0804}\ben
&&\eta(\Delta)=\op\sum_{0\leq|\La|\leq m}
t^r_\La\eta(\Delta)^{a,\La}_r
 w_a \in \cS^0_\infty[T,W^*], \label{0804}\\
&&\eta(\Delta)^{a,\La}_r=\op\sum_{0\leq|\Si|\leq
m-|\La|}(-1)^{|\Si+\La|}C^{|\Si|}_{|\Si+\La|} d_\Si
(\Delta^{a,\Si+\La}_r), \nonumber
\een
linear in $t^r_\La$ and $w_a$.
\end{theo}

\begin{proof}
The graded function $\up$ (\ref{0801}) defines the graded density
\be
\ol\up=\op\sum_{0\leq|\La|\leq m}
t^r_\La\up^{a,\La}_r(x^\la,s^A_\Si)
 \ol w_a\om\in \cS^0_\infty[T,\ol W^*].
\ee
Its Euler--Lagrange operator $\dl(\ol\up)$ (\ref{0709}) contains
the summand
\be
\cE_r\th^r\w\om=\op\sum_{0\leq|\La|\leq m} (-1)^{|\La|}\th^r\w
d_\La(\up^{a,\La}_r \ol w_a)\om,
\ee
which defines the graded density
\be
\op\sum_{0\leq|\La|\leq m} (-1)^{|\La|}t^r d_\La(\up^{a,\La}_r \ol
w_a)\om
\ee
owing to the canonical isomorphism $V^*T=T^*\op\times T$. Using
the relation (\ref{0606b}), we come to the formula (\ref{0802}).
The converse is proved similarly to the proof of Theorem
\ref{0610}.
\end{proof}

\end{document}